# Neutron-rich chromium isotope anomalies in supernova nanoparticles

N. Dauphas[1,2], L. Remusat[2,3], J.H. Chen[4,6], M. Roskosz[5,6], D.A. Papanastassiou[4,2], J. Stodolna[5], Y. Guan[2], C. Ma[2], J.M. Eiler[2]

[1]Origins Laboratory, Department of the Geophysical Sciences and Enrico Fermi Institute, The University of Chicago, 5734 South Ellis Avenue, Chicago, IL 60637, USA.

[2]Division of Geological and Planetary Sciences, California Institute of Technology, Pasadena, CA 91125, USA.

[3]Muséum National d'Histoire Naturelle, Laboratoire de Minéralogie et Cosmochimie du Muséum, CNRS UMR 7202, Case Postale 52, 61 rue Buffon, 75231 Paris, France.

[4]Jet Propulsion Laboratory, 4800 Oak Grove Drive, Pasadena, CA 91109.

[5]Unité Matériaux et Transformations, Université de Lille 1, CNRS UMR 8207, 59655 Villeneuve d'Ascq, France.

[6]These authors contributed equally to this work.

Correspondence should be addressed to N.D. (dauphas@uchicago.edu)

**ABSTRACT**

Neutron-rich isotopes with masses near that of iron are produced in type Ia and II supernovae. Traces of such nucleosynthesis are found in primitive meteorites in the form of variations in the isotopic abundance of $^{54}$Cr, the most neutron-rich stable isotope of chromium. The hosts of these isotopic anomalies must be presolar grains that condensed in the outflows of supernovae, offering the opportunity to study the nucleosynthesis of iron-peak nuclei in ways that complement spectroscopic observations and can inform models of stellar evolution. However, despite almost two decades of extensive search, the carrier of $^{54}$Cr anomalies is still unknown, presumably because it is fine-grained and is chemically labile. Here we identify in the primitive meteorite Orgueil the carrier of $^{54}$Cr-anomalies as nanoparticles, most likely spinels that show large enrichments in $^{54}$Cr relative to solar composition ($^{54}$Cr/$^{52}$Cr ratio >3.6×solar). Such large enrichments in $^{54}$Cr can only be produced in supernovae. The mineralogy of the grains supports condensation in the O/Ne-O/C zones of a type II supernova, although a type Ia origin cannot be excluded. We suggest that planetary materials incorporated different amounts of these nanoparticles, possibly due to late injection by a nearby supernova that also delivered $^{26}$Al and $^{60}$Fe to the solar system. This idea explains why the relative abundance of $^{54}$Cr and other neutron-rich isotopes vary between planets and meteorites. We anticipate that future isotopic studies of the grains identified here will shed new light on the birth of the solar system and the conditions in supernovae.

**1. INTRODUCTION**

All rocks on Earth have Cr isotopic compositions (relative abundances of $^{50}$Cr, $^{52}$Cr, $^{53}$Cr, and $^{54}$Cr) that very closely follow the laws of mass dependent fractionation (Ellis et al. 2002; Schoenberg et al. 2008). In contrast, some meteorites exhibit anomalous enrichments or depletions in the abundance of the neutron-rich isotope $^{54}$Cr, relative to the abundance expected based on co-existing Cr isotopes and the laws of mass-dependent fractionation (Birck & Allègre 1984; Papanastassiou 1986; Rotaru et al. 1992; Podosek et al. 1997, 1999; Nichols et al. 1998, 2000; Alexander 2002; Shukolyukov & Lugmair 2006; Trinquier et al. 2007; Birck et al. 2009; Qin et al. 2010). These anomalies most likely have a nucleosynthetic origin, meaning that they were produced by nuclear reactions in stars that existed before the Sun was born. They were presumably incorporated in solar system materials in the form of presolar grains that escaped volatilization in the protosolar nebula (Mendybaev et al. 2002). Presolar grains directly probe the composition of the astrophysical environment where they were condensed, and provide a view of stellar nucleosynthesis that complements information offered by astronomical techniques. Several types of presolar grains from various types of stars have already been identified in meteorites (Zinner 1998; Lodders & Amari 2005) but the carrier of $^{54}$Cr anomalies has eluded characterization. Previous measurements of bulk meteorite residues have shown that the carrier of $^{54}$Cr-anomalies must be smaller than 300 nm (Nichols et al. 1998, 2000). Several conference abstracts have reported the detection of $^{54}$Cr excess in individual grains from meteorites but many questions are left unanswered as to the exact nature of these grains (Ott et al. 1997; Qin et al. 2009; Dauphas et al. 2010; Nittler et al. 2010). To address the question of the nature and origin of the carrier of $^{54}$Cr anomalies, we have studied the Cr isotopic composition of residues from primitive meteorites using a high spatial resolution ion probe (Cameca NanoSIMS-50L).

## 2. MATERIALS AND METHODS
### 2.1. Preparation of Orgueil and Murchison residues

Based on previous studies (Rotaru et al. 1992; Podosek et al. 1997; Nichols et al. 1998, 2000), a protocol was established to concentrate the carrier of $^{54}$Cr-anomalies (Fig. 1). Approximately 1.4 g each of the meteorites Orgueil and Murchison were disaggregated at the University of Chicago using the freeze-thaw technique. The disaggregated meteorites were first treated with 5 mL of 50 % acetic acid (~9 mol/L) for 2 days at room temperature (~20 °C). The residues were then leached with 10 mL of 4.5 mol/L nitric acid for 10 days at ~20 °C. After removing the supernatants, solutions of 30-40 % sodium hydroxide (~13 mol/L) were added to the residues. Magnetic minerals that adhered to a Teflon coated magnetic stirring bar were removed. By rinsing with water, the pH decreased towards neutrality and large colloidal fractions were produced. The colloidal fractions were pipetted out and were then flocculated by acidification with 2 mol/L nitric acid, giving two residues denoted *Ocolloid* and *Mcolloid*. The fractions that readily settled at near-neutral pH (non-colloidal) were further treated with 2 mL of 8 mol/L nitric acid for 1 hour at ~70 °C followed by a day at ~20 °C. The residues were transferred in solutions of 0.5 % sodium hydroxide (~0.2 mol/L). Three size fractions were separated in that medium by differential sedimentation rates (Amari et al. 1994). The nominal size fractions calculated from Stokes' law assuming a density of 3 g/cm$^3$ for the grains are <200 nm (*O<200* and *M<200*), 200-800 nm (*O200-800* and *M200-800*), and >800 nm (*O>800* and *M>800*). The size distributions measured by Secondary Electron Microscopy (SEM) and Transmission Electron Microscopy (TEM) are in good agreement with calculated nominal sizes.

### 2.2. Transmission Electron Microscopy

Dried residues were suspended in ethanol by ultra-sonication and were then deposited as droplets on carbon-coated copper grids. The samples were studied at the University of Lille using a Tecnai G2-20 twin (LaB$_6$ filament, 200 kV). The microscope is equipped with an EDAX Si-detector for Energy Dispersive X-ray Spectroscopy (EDS). Grain microstructures were studied using bright field imaging in conventional TEM mode, selected area electron diffraction, and annular-bright field imaging in scanning (STEM) mode. Chemical compositions were studied by EDS in STEM configuration with probe sizes ranging from 5 to 10 nm. Chemical compositions were determined by standard deconvolution of the EDS spectra. The microstructure of the samples made it difficult to disentangle contributions from the organic matrix and the mineral grains. Furthermore, contamination of the samples (mainly Na and Cl from the chemical reagents used to prepare the residues) prevented direct quantification. Consequently, the analysis of a mineral grain was generally coupled to the analysis of the surrounding matrix for accurate background subtraction. A second method was to deconvolve the EDS spectra taking into account the contributions of all elements detected (including contaminants), but using only the elements expected in the mineral grain for the quantification. Calculations of element concentrations and atomic ratios were carried out using calibrated k-factors and thin film matrix correction procedures. The k-factors for the major elements were determined using standard specimens according to the parameter-less method of Van Cappellen (1990). The absorption correction procedure based on the principle of electroneutrality (Van Cappellen & Doukhan 1994) was applied whenever possible. Elemental distributions were obtained using spectral imaging wherein each pixel of an image contains a full EDS spectrum (see Leroux et al. 2008 for details). The uncertainty on atomic concentrations is typically in the range 5-15 %. Chromium concentrations as low as 200 ppm were previously quantified on the same microscope (Leroux et al. 2008). The minimum Cr concentration that could be reliably quantified in this study was ~100 ppm with a relative uncertainty of ±50 %.

### 2.2.1. Microstructure and mineralogy of the colloidal fraction

Samples from both Orgueil and Murchison were studied and the results presented below are the same for the two meteorites. The colloidal fraction is mainly composed of amorphous organic matter (OM). Local EDS analysis showed that this OM must contain less than ~50 ppm of chromium (signal below detection limit). Small crystalline grains are embedded in the fluffy OM of both Murchison and Orgueil separates (Fig. 2, 3). A large population of high-temperature minerals less than 50 nm in size is present in the matrix, namely SiC, anatase ($TiO_2$), and corundum ($Al_2O_3$). Small Fe-Ni metal grains (mainly in the size range 10-20 nm) were also observed. No chromium sulfide or phosphide grains, which had been suggested as possible carriers of $^{54}$Cr anomalies (Podosek et al., 1997), were found. Silica polymorphs (cristobalite) containing significant amounts of aluminum were found in Murchison and Orgueil. They generally occur as 50 nm-long crystals embedded in the matrix but were also occasionally found as larger chips (300 nm) in Murchison. Only one Cr-bearing diopside crystal was found among 150 analyses of crystalline grains. This crystal was found in Murchison and no such pyroxene was found in Orgueil. The dominant crystalline phase in our colloidal residues is spinel containing relatively large amounts of chromium (Table 1). In the colloidal fraction, these grains are always smaller than ~60 nm (Fig. 2A-D). Compositions and diffraction patterns of these grains are consistent with spinel group crystallography and chemistry. X-ray mapping of large clumps of OM and crystalline particles were performed to estimate the volume fraction of these spinels (Fig. 3). Conventional image processing performed on 4 different clumps of a few hundreds of cubic micrometers indicated a reproducible volume fraction of 900±100 ppm spinel in both Orgueil and Murchison. The composition of these minerals is highly variable (Table 1, Fig. 4). They are generally Cr-bearing spinels close to the Mg,Al-rich end-member of the group (i.e., $MgAl_2O_4$ with some Cr substituting for Al) but magnesiochromite ($MgCr_2O_4$) and chromite ($FeCr_2O_4$) are also present. The average Cr-concentration of 29 spinels less than 100 nm in size is 25.7 atomic% (normalized to 100 % non-oxygen atoms). Using these values, one can estimate the average Cr-content of bulk clumps of OM and crystalline particles such as the one showed in Fig. 3 to be 200±100 ppm by weight. This value is similar to the bulk value obtained by bulk EDS analysis. Though indirect, this mass balance calculation supports the idea that the only significant carriers of chromium are nanospinels.

### 2.2.2. Microstructure and mineralogy of the <200 nm size fraction

Compared to the colloidal fraction, the <200 nm nominal size fraction is composed of larger spinel grains and less organic matter. The grains are either completely separated from the OM or are stuck to small particles of OM (Fig. 2E, F). As expected, the grains are 150 to 200 nm in size and rarely exceed 300 nm. Contrary to the large compositional variations found in the nanospinels of the colloidal fraction, the composition of the large spinels is much more homogeneous (Fig. 4). The Al-content is extremely low and most of them can be described as pure chromite grains. Eskolaite ($Cr_2O_3$) is also common in this fraction.

### 2.2.3. Effect of HCl leaching on the studied objects

Samples of the colloidal and <200 nm size fractions were treated with 5 mol/L HCl for 1 day at 80 °C. Previous work has shown that such treatment releases Cr from phases that have large $^{54}Cr$ excess. In the colloidal fraction after treatment with HCl, no spinel was found in small particles of OM. Only C, S, O, and contaminants like Na and Cl were detected but no Cr was measured. However, we occasionally found some small eskolaite grains, not necessarily associated with the organic matrix, which may be relicts of incompletely dissolved coarser grains. The spatial density of grains in the leached fraction is much smaller compared to that before leaching, indicating that the grains were dissolved by hot HCl (Fig. 5).

In large particles of OM (several micrometers in length), small spinels and even metal nuggets are still present in the innermost part, while they are absent in a leached layer in the outer part. Thus, small spinels seem to be readily dissolved by HCl, unless they are embedded in a protective layer of OM. This point is further supported by TEM study of the <200 nm size fraction. After the same HCl treatment, some spinels and eskolaite are recovered but their size is only 50-100 nm compared to 150-200 nm before leaching. This size reduction must owe to the fact that they have been partially dissolved by hot HCl.

**2.3. Thermal ionization mass spectrometry**

2.3.1. Sample dissolution and chemical separation

Both the Murchison and Orgueil residues were dissolved in 1 mL of concentrated HF (~32 mol/L) and 0.1 mL of concentrated $HNO_3$ (~16 mol/L) using a small, high P-T Teflon capsule at 180 °C for 4 days. After dissolution, the solution was evaporated to dryness. Then 1 mL of concentrated HCl (~12 mol/L) and 0.1 mL of concentrated $HNO_3$ were added to the residue (in the Teflon capsule) and heated again at 180 °C in an oven for 2 days. The HCl solution was placed in a micro-centrifuge tube and was centrifuged at 1600 rpm (~250 $g$) for 1 minute. Usually, this will produce a clear solution with little or no residue even if the original sample contains graphite. If a residue was found after the high P-T treatment (only the "colloidal" samples produced a small amount of residue, < 1 mg), the above HF+$HNO_3$ and HCl+$HNO_3$ dissolution steps were repeated 2-3 times (at 200-220 °C) until no residue was found or the residue remained constant. Under these conditions, most if not all spinel is dissolved.

After dissolution of the sample, the solution was converted to the chloride form (in 6-9 M HCl). A small aliquot was taken to determine the Mn and Cr contents using a graphite furnace atomic absorption spectrometer. Another aliquot was taken for Cr isotopic analysis. Using a 2-ml anion-exchange column (column 1, AG1W-X8) Cr is readily separated (not adsorbed) from other major elements (Fe, Cu, Co and Ti) in 6-9 M HCl. The separation of Cr from the matrix elements makes use of the slow re-equilibration of the different hydrated species of Cr(III) in aqueous solutions (Birck & Allègre, 1988). Following the procedure developed by Chi-Yu Shih (personal communication, 2009), Cr in solution was converted to Cr(III) by reduction in 6 M HCl at 150 °C for 24 hours. This solution was evaporated to near dryness followed by dilution with 0.5 ml 1 M HCl and 0.5 ml water, which was loaded onto a 2-ml cation exchanger (column 2, AG 50W-X8 200-400 mesh resin). Chromium was eluted in the first 6-7 ml of 1 M HCl. The above step was repeated if there was a noticeable amount of impurity (e.g., Ca, Al and Mg measured by atomic absorption) in the Cr solution. If traces of iron were detected in the Cr solution, the solution was evaporated to near dryness and processed through anion-exchange (AG1W-X8) using a smaller (0.1 ml) column. Occasionally when processing a larger sample (> 50 mg), a second cation exchange column (column 3, 0.33 ml, AG50W-x8 200-400 mesh) as described by Trinquier et al. (2008) was used. The recovery of Cr was ~100 % for column 1, ~80 % for column 2 and ~100% for column 3. The sample was then ready for mass spectrometric analysis. The total chemistry blanks of 5 ng are negligible compared to the amounts of Cr (3 to 37 μg) extracted from each sample.

### 2.3.2. Mass spectrometry

An aliquot of 1-3 μg of the clean Cr sample was evaporated to dryness under UV light to eliminate any organic residue. The Cr was dissolved in a small drop of dilute $HNO_3$ and loaded with a mixture of silica gel, aluminum and boric acid onto a Re, V-shaped filament. The Cr isotopic composition was determined using a Thermo Scientific Triton Thermal Ionization Mass Spectrometer in a dynamic mode (Table 2). We did not use the L1 and H1 cups because they are not optimal for the Cr mass region. Mass-dependent isotope fractionation in the $^{53}Cr/^{52}Cr$ and $^{54}Cr/^{52}Cr$ ratios (line 1) were corrected by internal normalization to the $^{50}Cr/^{52}Cr$ (line 2) using a value of 0.051859 (Shields et al. 1966) and the exponential law (Russell et al., 1978). The interfering isotopes were monitored both on-line (real-time) with data collection and before and after each run using an electron multiplier mass scan. Interferences from $^{54}Fe$, $^{50}Ti$ and $^{50}V$, monitored at masses 56, 48, and 51, were always negligible (normal isotopic composition was assumed for the interfering elements). Chromium standards (NIST SRM 3112a) were measured before and after each Cr sample. The Cr data were collected over two intervals in August and October 2009. We replaced a pre-amplifier for the center cup in September 2009, which resulted in a small shift in the Cr isotopic ratios between the two sessions.

After normalization, the means and standard deviations ($2\sigma_{mean}$) of each run are calculated, with elimination of outliers (fewer than 5% of the measured cycles). The uncertainties reported for each analysis represent a total of 600-1200 ratios. This quoted uncertainty is the internal precision of a single analysis, with the data obtained under similar conditions (e.g., ion signal intensity, range of mass-dependent isotope fractionation, absence of mass interferences). The internal precision is used as a measure of the quality of individual analyses. For comparing samples, the external precision is used and is defined by the reproducibility of the analyses of the Cr standard. The external relative precisions (1σ) are 5-6 and 21-64 ppm for $^{53}Cr/^{52}Cr$ and $^{54}Cr/^{52}Cr$ ratios, respectively. The deviations and errors of the $^{53}Cr/^{52}Cr$ and $^{54}Cr/^{52}Cr$ ratios in the samples are expressed as ε values relative to the mean value of the standard during each session, where $\varepsilon^{i}Cr=[(^{i}Cr/^{52}Cr)_{sample}/(^{i}Cr/^{52}Cr)_{std}-1]\times10^4$.

## 2.4. NanoSIMS measurements and data reduction
### 2.4.1. Instrumental setting and data acquisition

NanoSIMS imaging was performed on the Cameca NanoSIMS 50L installed at the Caltech Center for Microanalysis during three sessions covering a total duration of 24 days. A primary O⁻ beam of ~30 pA, with an energy of 16 kV, was rastered on surface areas of 20×20 or 40×40 μm² at a raster speed of 20 ms/pixel. The spatial resolution with the O⁻ beam was between 400 and 800 nm. The images were acquired with resolutions of 256×256 and 512×512 pixels for 20×20 and 40×40 μm² raster areas, respectively (78 nm/pixel). Secondary ion images of $^{52}$Cr$^+$, $^{53}$Cr$^+$, $^{54}$Cr$^+$+$^{54}$Fe$^+$, $^{56}$Fe$^+$, and $^{57}$Fe$^+$ were recorded in multi-collection mode, using Hamamatsu discrete dynode electron multipliers with a dead time of 44 ns. Before each analysis, a high primary current (around 600 pA) was used to pre-sputter the surface to achieve a steady state of secondary ion emission and to remove surface contamination. Vacuum in the analysis chamber was ~2.7×10$^{-9}$ mbar. The contribution of hydride molecular ions was negligible. Terrestrial chromite was analyzed everyday to check stability, reproducibility and to correct the data. Chromium carbide and magnetite were also used in each session to refine and validate the correction of $^{54}$Fe interference on $^{54}$Cr. The useful yield of Cr was ~3.7 times that of Fe.

### 2.4.2. Data reduction

The mass resolving power (MRP=M/ΔM) was about 9,000 on $^{52}$Cr, which is insufficient to resolve the interference of $^{54}$Fe on $^{54}$Cr (requiring a MRP of >73,000). The contribution of $^{54}$Fe$^+$ on $^{54}$Cr$^+$ was therefore corrected using $^{56}$Fe$^+$ intensity and a fixed $^{54}$Fe/$^{56}$Fe ratio,

$$\frac{^{54}Cr}{^{52}Cr} = \frac{^{54}i^+}{^{52}i^+} - \frac{^{56}i^+}{^{52}i^+} \times \left(\frac{^{54}Fe}{^{56}Fe}\right)_{terrestrial}.$$

The correction was adjusted using magnetite measurements by modifying the assumed (terrestrial) $^{54}$Fe/$^{56}$Fe ratio. The $^{57}$Fe/$^{56}$Fe ratio was monitored to make sure that the assumption of uniform Fe isotopic composition was correct. Mass fractionation on Cr isotopes was either not corrected or was corrected by internal normalization assuming a terrestrial value for the $^{53}$Cr/$^{52}$Cr ratio (Shields et al., 1966) and using the exponential law (Russell et al. 1978).

Image analysis was performed following two different approaches. In method 1, regions of interest (ROIs) corresponding to isolated grains were defined using a grain identification algorithm included in *L'image* software (L.R. Nittler, Carnegie Institution of Washington). For each ROI, the total counts were retrieved to calculate $^{54}$Cr/$^{52}$Cr isotopic ratios as previously described. The drawback of this approach is that one may miss anomalies carried by small or Cr-poor grains. In method 2, corrected $^{54}$Cr/$^{52}$Cr isotope ratio images were generated using a *Mathematica* routine and *L'image* software. These images were used to locate $^{54}$Cr-rich "hot-spots". Corrected $^{54}$Cr/$^{52}$Cr ratios were calculated from the total counts in each ROI corresponding to these hot spots. To generate isotopic ratio images, the isotopic ratio of each pixel was normalized to the isotopic ratio of the bulk image. This approach allows us to identify isotopic anomalies that do not relate to any resolvable grain. Isotopic ratio images generated using *Mathematica* and *L'image* software are almost identical (compare Fig. 6A, B).

### 2.4.3. Assessment of uncertainties

In order to assess the total uncertainty on $^{54}Cr/^{52}Cr$ ratios, ion images of isotopically and chemically homogeneous chromite grains were subdivided into grids of squares or hexagons of uniform sizes. For example, a 10×10 μm² image of 128×128 pixels was divided into 9 ROIs of 32×32 pixels each, 49 ROIs of 16×16 pixels, 225 ROIs of 8×8 pixels, 961 ROIs of 4×4 pixels, and 4096 ROIs of 2×2 pixels. The number of pixels within all ROIs does not always add up to the number of pixels in the total image because ROIs cannot always be defined near the borders. For each unit grid size, the standard deviation of the isotopic ratios of all ROIs was compared with that predicted from counting statistics. For $^{53}Cr/^{52}Cr$ and $^{57}Fe/^{56}Fe$ ratios, the measured dispersion is entirely accounted for by total ion counts and Poisson statistics. The $(^{54}Cr+^{54}Fe)/^{52}Cr$ and $^{56}Fe/^{52}Cr$ ratios vary more than what is allowed by counting statistics, presumably because of additional variability during secondary ion formation of Cr and Fe. The uncertainties on these ratios can be expressed as the sum of the uncertainty associated with counting statistics and a term that is proportional to the $^{56}Fe^+/^{52}Cr^+$ ratio:

$$\sigma_{53/52} = \sigma_{Poisson}$$
$$\sigma_{54/52} = 0.0027 \times \left(56^+/52^+\right) + \sigma_{Poisson}$$
$$\sigma_{56/52} = 0.053 \times \left(56^+/52^+\right) + \sigma_{Poisson}$$
$$\sigma_{57/56} = \sigma_{Poisson}$$

The factors of 0.0027 and 0.053 were obtained by plotting the dispersion in the measured ratios of the standard chromite grains ($\sigma_{54/52}$ and $\sigma_{56/52}$) as a function of the error associated with counting statistics ($\sigma_{Poisson}$). For $(^{54}Cr+^{54}Fe)/^{52}Cr$ and $^{56}Fe/^{52}Cr$ ratios, a linear relationship was found with a *y*-intercept different from zero. The *y*-intercept was assumed to scale linearly with the measured $(56^+/52^+)$ ratio. These relationships were used to calculate uncertainties on isotopic ratios from total counts in ROIs. Error propagation was used to calculate the uncertainty on the $^{54}Cr/^{52}Cr$ ratio corrected for $^{54}Fe$ isobaric interference:

$$\sigma_{^{54}Cr/^{52}Cr} = \sqrt{\sigma_{54/52}^2 + 0.0637^2 \times \sigma_{56/52}^2}.$$

Another approach was adopted for assessing uncertainties on isotope ratio images and for evaluating the significance of $^{54}Cr$-rich regions. Ion count images of $^{52}Cr$ and $^{56}Fe$ were used to generate synthetic images of $^{53}Cr$ and $^{54}Cr+^{54}Fe$ using terrestrial $^{53}Cr/^{52}Cr$, $^{54}Cr/^{52}Cr$, and $^{54}Fe/^{56}Fe$ ratios. By construction, these images have normal (terrestrial) isotopic compositions and have a spatial distribution of $^{52}Cr$ ion counts that is identical to that of the real image. A random number generator following Poisson distribution is then applied to each pixel to simulate the presence of noise associated with limited counting statistics. Those images are then subjected to the same treatment as that applied to real data. A *Mathematica* script modified from the one written to analyze real data was used to generate these synthetic images. The synthetic maps of normal (terrestrial) $^{54}$Cr/$^{52}$Cr ratio with noise added can then be compared directly with measured isotope ratio images (compare Fig. 6B, C).

## 3. RESULTS

Approximately 1.4 g each of the carbonaceous chondrites Orgueil (classified as CI1) and Murchison (CM2) were treated by a sequence of acid leaching, magnetic separation, and size separation (Nichols et al. 1998, 2000) (Fig. 1). Because leaching with hot HCl is known to partially dissolve the carrier of $^{54}$Cr anomalies, this acid was avoided during preparation of the residues. For each meteorite, four residues differing in nominal grain size were isolated (colloidal, <200 nm, 200-800 nm, and >800 nm). Bulk Cr isotopic compositions of all residues were measured by Thermal Ionization Mass Spectrometry at the Jet Propulsion Laboratory (Table 3, Fig. 7). All four residues from each meteorite show a large excess $^{54}$Cr. The colloidal fractions present the largest anomalies, with $\varepsilon^{54}$Cr values $\{\varepsilon^{54}\text{Cr}=[(^{54}\text{Cr}/^{52}\text{Cr})_{sample}/(^{54}\text{Cr}/^{52}\text{Cr})_{standard} -1]\times10^4\}$ of +170.26±0.11 and +148.03±0.17 in Orgueil and Murchison, respectively. This is much higher than previously reported values of $\varepsilon^{54}$Cr~10 to 20 in bulk residues of carbonaceous chondrites (Shukolyukov & Lugmair 2006; Qin et al. 2010).

Individual Cr-rich grains in the colloidal fraction are too small to be analyzed by NanoSIMS (~30 nm median size, measured at Université de Lille 1 by TEM). We therefore focused on larger grains, in particular those from the <200 nm nominal size fraction of Orgueil (bulk $\varepsilon^{54}$Cr=+38.60±0.12). A suspension of these grains was deposited on a gold foil and the Cr isotopic composition of the grains was mapped using a NanoSIMS-50L. The interference on $^{54}$Cr$^+$ from $^{54}$Fe$^+$ was corrected by measuring $^{56}$Fe$^+$ ion beam intensity and subtracting $^{54}$Fe$^+$ calculated using $^{56}$Fe$^+$ and assuming a known (terrestrial) $^{54}$Fe$^+$/$^{56}$Fe$^+$ ratio. The total mapped surface area was ~31,000 $\mu$m$^2$ (equivalent to a square of 175×175 $\mu$m) from which 1,437 regions of interest (ROIs) were isolated. All ROIs, except two, have normal (terrestrial) $^{54}$Cr/$^{52}$Cr ratios within error (Fig. 8). The two anomalous grains denoted np1 and np2 hereafter (for nanoparticles 1 and 2), have $^{54}$Cr/$^{52}$Cr ratios of 3.59(±0.29)×solar ($\varepsilon^{54}$Cr=+25,861±2,903) and 1.27(±0.09)×solar ($\varepsilon^{54}$Cr=+2,654±915), respectively (Table 4, Fig. 9, 10, 11). The same spot where np1 was found was measured a second time on a different day and the $^{54}$Cr excess was confirmed ($^{54}$Cr/$^{52}$Cr=2.79×solar; $\varepsilon^{54}$Cr=+17,902±3,291). The different compositions are explained by different dilution with normal Cr from adjacent grains.

In np1, the $^{54}$Fe correction on $\varepsilon^{54}$Cr of -5,775 (corresponding to a Fe/Cr ratio of ~0.7) is small compared to the measured anomaly of +25,861. This correction is calculated assuming constant $^{54}$Fe/$^{56}$Fe ratio (terrestrial), which may not be valid if the grain formed in a supernova. A $^{54}$Fe/$^{56}$Fe ratio of ~2×solar can be produced in type Ia supernovae (Woosley 1997), which would increase the $^{54}$Fe interference correction and would bring the $^{54}$Cr/$^{52}$Cr ratio of np1 to 3.0×solar. A $^{54}$Fe/$^{56}$Fe ratio between ~0.01 and 0.7×solar can be produced in the O/Ne and O/C zones of type II supernovae (Rauscher et al. 2002; http://nucastro.org/nucleosynthesis/), which would reduce the $^{54}$Fe interference correction and would bring the $^{54}$Cr/$^{52}$Cr ratio of np1 to 4.2×solar. However, the measured $^{57}$Fe/$^{56}$Fe ratio of np1 is terrestrial, suggesting that most Fe in that ROI is from adjacent isotopically normal grains. Thus, it is likely that the $^{54}$Fe correction assuming solar $^{54}$Fe/$^{56}$Fe ratio is accurate. To summarize, the anomalies correspond to real excess $^{54}$Cr and not inadequate correction of $^{54}$Fe interference.

## 4. $^{54}$Cr-RICH SUPERNOVA NANOPARTICLES IN PLANETARY MATERIALS

Assuming that all grains contain the same amount of Cr and given that ~1 grain out of 1,000 contains anomalous $\varepsilon^{54}$Cr=+26,000, one would predict that the bulk residue of Orgueil <200 nm size fraction should have $\varepsilon^{54}$Cr~+26, which is close to the measured value of +38.60. Thus, the $^{54}$Cr-rich grains identified in this study are likely to be the sole carriers of the $^{54}$Cr anomalies measured in this sample and perhaps in planetary materials, in general (Rotaru et al. 1992; Podosek et al. 1997, 1999; Nichols et al. 1998, 2000; Alexander 2002; Shukolyukov & Lugmair 2006; Trinquier et al. 2007; Birck et al. 2009; Qin et al. 2010). These grains have terrestrial isotopically normal $^{53}$Cr/$^{52}$Cr ratios, which is consistent with the bulk $^{53}$Cr/$^{52}$Cr ratio of the <200 nm nominal size fraction of Orgueil being close to terrestrial ($\varepsilon^{53}$Cr=-0.66±0.05).

The spatial resolution of the NanoSIMS O$^-$ primary beam does not allow us to precisely determine the size of grains smaller than ~600 nm in diameter. Under a Field Emission Scanning Electron Microscope (FE-SEM), several grain-like structures, all less than ~100 nm in size, were found at the location of np1, but the grain itself could not be identified. This is consistent with the fact that the largest anomalies in the bulk residues were found in the colloidal fraction that contains grains that are less than ~100 nm in size (Fig. 7). In bulk, the colloidal fractions have $\varepsilon^{54}$Cr values (+170 and +148 for Orgueil and Murchison, respectively) that approach those found previously in acid leachates, indicating that the carrier of $^{54}$Cr-anomalies was efficiently concentrated in those fractions. Therefore, we performed a detailed TEM study of the colloidal residue to identify Cr-bearing phases. The dominant crystalline phases are nano-spinels containing relatively large amounts of chromium [*e.g.*, $Mg(Al,Cr)_2O_4$]. These spinels are embedded in organic matter and are sometimes associated with carbon nano-globules (Fig. 2D), which commonly have high $^{15}$N/$^{14}$N and D/H ratios indicative of low temperature chemistry in cold molecular cloud or outer protosolar disk environments (Nakamura-Messenger et al. 2006). It is thus likely that the carrier of $^{54}$Cr-anomalies is an oxide grain, and likely a spinel. Most nano-spinels have solar composition and it is presently unknown if anything distinguishes mineralogically or chemically the carrier of $^{54}$Cr-anomalies from similar grains formed in the solar system.

To assess the chemical resistance of chromium-bearing nano-spinels, we have subjected the residues to 5 mol/L HCl for a day at 80 °C. The residues were then reexamined under a TEM. In the colloidal fraction, the nano-spinels survived when they were embedded in compact organic matter but they disappeared from the more porous fluffy aggregates (Fig. 5). In the <200 nm nominal size fraction of Orgueil, the median of the size distribution shifted from 184 to ~80 nm, indicating that they were partially digested. It appears that hot HCl can dissolve spinels that are smaller than 100 nm in a relatively short time.

Because of insufficient spatial resolution of the NanoSIMS, the signal from the $^{54}$Cr-rich grains may have been diluted by normal chromium isotopic composition from surrounding grains. Thus, the true $^{54}$Cr/$^{52}$Cr ratio of the carrier phase may be higher than the values reported here and the 3.6×solar enrichment in the $^{54}$Cr/$^{52}$Cr ratio of np1 represents a well-resolved anomaly but a lower limit to the actual ratio. Such a large enrichment cannot be produced in the solar system and must have a stellar origin. The Cr isotopic compositions of some presolar grains (SiC and spinels) thought to have condensed in the outflows of AGB-stars have been measured and the maximum enrichment in the $^{54}$Cr/$^{52}$Cr ratio is ~1.1×solar, in agreement with predictions for AGB-star nucleosynthesis (Zinner et al. 2005; Levine et al. 2009; Savina et al. 2010). Therefore, AGB-stars can be ruled out as the source of $^{54}$Cr-anomalies in solar system materials. Chromium-54 can be produced in both SNIa (thermonuclear explosion of a carbon-oxygen white dwarf) and SNII (core-collapse and explosion of a massive star) supernovae. It is estimated that approximately 2/3 of the solar system $^{50}$Ti came from SNIa and the other 1/3 was produced in SNII (Meyer et al. 1996). The same proportion also applies to $^{54}$Cr (B.S. Meyer, personal communication).

The nucleosynthesis of $^{54}$Cr holds clues on the origin and evolution of type Ia supernovae (Woosley 1997; Brachwitz et al. 2000; Thielemann et al. 2004). SNIa are formed in binary systems, when a white dwarf accretes matter from a companion star and carbon fusion is ignited as the star approaches the Chandrasekhar limit. In the inner part of the SNIa, iron-peak nuclei are produced by nuclear quasi-equilibrium. If the central density at ignition is high enough, electron capture can occur, which produces neutron-rich isotopes like $^{48}$Ca or $^{54}$Cr (high central ignition densities are the result of low accretion rates on the white dwarf). In detail, the flame propagation speed also influences the production of $^{54}$Cr. In Fig. 12A, we compare the measured Cr isotopic composition of $^{54}$Cr-rich grain np1 with model predictions (slow deflagration models NCD2A-NCD8A of Woosley 1997) for a range of central ignition densities between $2\times10^9$ and $8\times10^9$ g cm$^{-3}$ (above ~ $9\times10^9$ g cm$^{-3}$, nuclear burning in the white dwarf induces collapse to a neutron-star rather than explosion as a supernova). The predicted $^{54}Cr/^{52}Cr$ ratios are probably too high because the electron capture rates used by Woosley (1997) have been subsequently revised downward (Brachwitz et al. 2000) but no comprehensive study of the nucleosynthesis of neutron-rich isotopes has been published with these reduced rates. As shown, the Cr isotopic composition of np1 ($^{54}Cr/^{52}Cr=3.6\times$solar) can be explained by nucleosynthesis in a SNIa using a central ignition density of $\sim2\times10^9$ g cm$^{-3}$. Owing to dilution by surrounding grains, the measured $^{54}Cr/^{52}Cr$ ratio represents a lower limit on the true ratio of the grain so higher ignition densities are also permitted by this constraint. Individual presolar grains, such as the ones identified in this study, record the composition of a particular parcel of matter in the star. Rather than comparing measured isotopic ratios with total yields, they could be compared with nucleosynthetic predictions for tracer particles in multi-dimensional simulations of SNIa explosions (Travaglio et al. 2004, 2005; Brown et al. 2005; Röpke et al. 2006; Meakin et al. 2009; Fink et al. 2010; Maeda et al. 2010). Such simulations are computationally challenging and the parameter space explored so far does not include SNIa progenitors capable of producing neutron-rich isotopes in large quantities.

In SNII, neutron-rich isotopes are produced by neutron capture reactions during core C- and shell He-burning (Rauscher et al. 2002; The et al. 2007; Pignatari et al. 2010). This is known as the weak *s*-process. Chromium-54 is made during the presupernova evolution of those stars by capture on $^{53}$Cr of neutrons produced by the $^{22}$Ne($\alpha$,n)$^{25}$Mg reaction. The other isotopes of Cr are produced as radioactive progenitors (*e.g.*, $^{52}$Fe for $^{52}$Cr) in the inner region of the SNII by nuclear quasi-equilibrium associated with explosive O and Si burning. Thus, the Cr isotopic composition in one-dimensional SNII models is concentrically zoned. In Fig. 12B, we compare the Cr isotopic ratios measured in np1 with SNII model predictions for a 21 M$_\odot$ progenitor star (Rauscher et al. 2002; http://nucastro.org/nucleosynthesis/). In the region of the star where neutron-capture was active (O/Ne and O/C zones, between ~2.5 and 5.3 M$_\odot$ in mass coordinate), the $^{54}Cr/^{52}Cr$ ratio can reach values as high as ~120×solar, which can explain the composition measured in the grain. Very similar enrichments are found for all SNII progenitor masses studied by Rauscher et al. (2002) (15, 19, 20, 21 and 25 M$_\odot$, http://nucastro.org/nucleosynthesis/).

Both type Ia and II supernovae can produce the large enrichments in $^{54}$Cr that have been measured in individual grains from meteorite residues. However, one might expect the dust that condensed in the outflows from these two types of stars to be different (Meyer et al. 1996; Woosley 1997). In SNII, the zones where the $^{54}$Cr-enrichment is the highest have O/C>1, so equilibrium thermodynamic calculations predict that dust produced in those regions should be dominated by oxide grains (Fedkin et al. 2010). For instance, it is predicted that spinels should condense as $MgAl_2O_4$ at ~1,600 K in O/C and O/Ne zones but that during expansion and cooling, ~20 mol% of $MgCr_2O_4$ would dissolve in the spinel at 1,000 K if equilibrium conditions are maintained. As discussed previously, the carrier of $^{54}$Cr-anomalies is likely to be a nanospinel, which would be consistent with condensation in the ejecta of a SNII. In SNIa, the place where $^{54}$Cr is produced does not contain oxygen. Yet, study of multi-epoch spectra of SNIa shows that mixing can take place (Stehle et al. 2005; Mazzali et al. 2008) and one cannot exclude the possibility that Cr-bearing oxides could condense in the outflows of those stars (either as new grains or as coatings on grains already present in the interstellar medium —ISM).

Measuring the isotopic abundance of $^{48}$Ca in nano-spinels will provide a definitive test to distinguish between SNIa and SNII origins, as high density SNIa represent the only conceivable stellar source for $^{48}$Ca (Meyer et al. 1996; Woosley 1997). Meteorite leachates do not show correlated isotopic anomalies for $^{48}$Ca and $^{54}$Cr (Moynier et al. 2010) but this could be a mineralogical effect as the nanospinels identified here do not contain appreciable amounts of Ca. FUN and hibonite-bearing inclusions in meteorites contain correlated excess in $^{48}$Ca and other neutron-rich isotopes that are consistent with a SNIa origin (Völkening & Papanastassiou 1989; Meyer & Zinner 2006). Future work will tell whether a link exists between these anomalies and the carrier of $^{54}$Cr-anomalies identified here.

### 5. COSMIC MEMORY OR INJECTION FROM A NEARBY SUPERNOVA?

Small, yet resolvable $^{54}$Cr anomalies are present in bulk planetary materials (Podosek et al. 1999; Shukolyukov and Lugmair 2006; Trinquier et al. 2007; Qin et al. 2010), which can be explained by large-scale heterogeneous distribution of the carrier of $^{54}$Cr anomalies in the protosolar nebula. Such heterogeneity could have been inherited from the molecular cloud core that made the solar system. The grains identified in this study could have been formed in supernovae that exploded long before the formation of the solar system. Even if these grains were homogeneously distributed in the ISM, isotopic anomalies at the scale of planetesimals could have been recovered by gas-dust decoupling (Adachi et al. 1976; Weidenschilling 1977). As shown in the present study, $^{54}$Cr-anomalies are carried by grains less than 100 nm that would follow the gas. Aerodynamic sorting of isotopically normal, coarser grains, could have produced large-scale isotopic anomalies for $^{54}$Cr. A correlation observed between $^{54}$Cr/$^{52}$Cr and Mn/Cr ratios among carbonaceous chondrites (Shukolyukov and Lugmair 2006; Qin et al. 2010) is consistent with this interpretation. However, no correlation was found with other proxies of volatile element depletion (Zn/Si ratio) or size sorting (matrix fraction) in meteorites (Qin et al. 2010). Chromium-54 seems to correlate with anomalies in $^{50}$Ti (Trinquier et al. 2009). This could arise from the fact that these two neutron-rich isotopes were produced at the same site and that the carriers have the same size and the same chemical properties, so that they would not be decoupled during incorporation in meteorites.

The presence of short-lived nuclides like $^{26}$Al (Lee et al. 1976, 1977; Jacobsen et al. 2008; MacPherson et al. 2010) and $^{60}$Fe (Birck & Lugmair 1988; Shukolyukov & Lugmair 1993; Tachibana & Huss 2003; Mostefaoui et al. 2005; Mishra et al. 2010) at the time of formation of meteorites supports birth of the solar system near a core-collapse supernova (Cameron & Truran 1977; Foster & Boss 1996; Vanhala & Boss 2000; Chevalier 2000; Ouellette et al. 2007). An appealing possibility is that the supernova that injected these extinct radioactivities into the early solar system also delivered the $^{54}$Cr-rich nanoparticles identified in the present study. A test of this hypothesis will be to search for correlations between $^{54}$Cr anomalies and the initial abundance of extinct radioactivities like $^{60}$Fe in meteorites, as this nuclide is produced together with $^{54}$Cr. Assuming that $^{54}$Cr-rich nanoparticles and $^{60}$Fe share the same core-collapse supernova (ccSN) origin, one can for example predict the degree to which the initial $^{60}$Fe/$^{56}$Fe ratio in meteorites should be sensitive to $^{54}$Cr/$^{52}$Cr isotopic variations using an equation similar to Eq. (C5) of Dauphas et al. (2008):

$$\left(^{60}Fe/^{56}Fe\right)_{r,t_0} = \left(^{60}Fe/^{56}Fe\right)_{CHUR,t_0} + \frac{10^{-4}}{c} \frac{1+\rho_{Fe}^{56}}{\rho_{Cr}^{54} - \mu_{Cr}^{54}\rho_{Cr}^{50}} \left(^{60}Fe/^{56}Fe\right)_{ccSN} e^{-\lambda_{60}\Delta t} \times \left(\varepsilon^{54}Cr_r - \varepsilon^{54}Cr_{CHUR}\right)$$

,
where $(^{60}Fe/^{56}Fe)_{r,t0}$ and $(^{60}Fe/^{56}Fe)_{CHUR,t0}$ are the ratios in reservoir $r$ and in CHUR (CHondritic Uniform Reservoir, taken here to be a reservoir with initial $^{60}Fe/^{56}Fe=6\times10^{-7}$ and $\varepsilon^{54}Cr=0$), $c=(^{52}Cr/^{54}Fe)_{ccSN}/(^{52}Cr/^{54}Fe)_{CHUR}$, $\rho_{Cr}^i=(^iCr/^{52}Cr)_{ccSN}/(^iCr/^{52}Cr)_{CHUR}-1$, $\rho_{Fe}^{56}=(^{56}Fe/^{54}Fe)_{ccSN}/(^{56}Fe/^{54}Fe)_{CHUR}-1$, $\mu_{Cr}^{54}=(54-52)/(50-52)$, and $\varepsilon^{54}Cr$ is as defined in Sect. 2.3.2 (deviation in part per 10,000 of the $^{54}$Cr/$^{52}$Cr ratio corrected by internal normalization to a constant $^{50}$Cr/$^{52}$Cr ratio relative to a reference composition). Note that this equation assumes that $^{54}$Cr and $^{60}$Fe were not chemically decoupled during injection in the solar system. For this calculation, we used the composition of the bulk O/Ne and O/C layers of a 21 M$_\odot$ progenitor star (Rauscher et al. 2002, http://nucastro.org/nucleosynthesis/), an initial $^{60}$Fe/$^{56}$Fe ratio of $6\times10^{-7}$ for CHUR (Mishra et al. 2010), and a free-decay interval of 1 My between nucleosynthesis in the supernova and injection in the solar system (Meyer & Clayton 2000). The half-life of $^{60}$Fe is that recently measured by Rugel et al. (2009) of 2.62 My. The prediction is that for the range of $\varepsilon^{54}$Cr values measured in bulk meteorites (from -0.8 in achondrites to +1.6 in CI chondrites; Trinquier et al. 2007; Qin et al. 2010), variations in the initial $^{60}$Fe/$^{56}$Fe ratio should be on the order of 10-20 % (Fig. 13). Such variations will be challenging to detect if present. The results obtained so far on the relative abundance of the neutron-rich isotope $^{58}$Fe provide no evidence for $^{60}$Fe heterogeneity at the scale of bulk planetary objects outside of ±10 % (Dauphas et al. 2008). FUN inclusions in meteorites show large variations in $\varepsilon^{54}$Cr, from -23 to +48 (Papanastassiou 1986). Variations of $^{48}$Ca in these inclusions clearly favor a SNIa origin (Meyer & Zinner 2006) but Kratz et al. (2001) argued that these anomalies could also be produced in SNII. If the $^{54}$Cr isotopic anomalies found in FUN inclusions derive from injection from a nearby supernova, we would predict that the initial $^{60}Fe/^{56}Fe$ ratio in these objects should vary between ~0 ($\varepsilon^{54}Cr$ =-23) and 26×10$^{-6}$ ($\varepsilon^{54}Cr$ =+48). While this represents a very large range, the main difficulty with these inclusions is their mineralogy, which is not suited to establish $^{60}Fe$ isochrons.

**6. SUMMARY AND CONCLUSIONS**

Meteorites and planets have variations in the abundance of the neutron-rich isotope $^{54}Cr$. The nature of the carrier of these anomalies is one of the key unsettled questions in cosmochemistry and planetary sciences. In order to address this question, we have measured the Cr isotopic composition of physical and chemical separates from the primitive carbonaceous chondrites Orgueil and Murchison (Fig. 1). The data have clear implications.

1. The colloidal fraction from Orgueil, which has a median grain size of ~30 nm, shows the largest $^{54}Cr$-excess ever measured in a bulk meteorite residue ($^{54}Cr/^{52}Cr$=1.017×solar, Fig. 7). This indicates that the carrier of $^{54}Cr$-anomalies was efficiently concentrated by our procedure and that it must be very fine grained (<100 nm).

2. In situ Cr isotopic analyses by secondary ion mass spectrometry of the <200 nm nominal size fraction of Orgueil (median grain size 184 nm) revealed the presence of $^{54}Cr$-rich nanoparticles ($^{54}Cr/^{52}Cr$>3.6×solar, Fig. 11). Mass balance shows that heterogeneous distribution of these nanoparticles in the inner solar system can explain the variations in $^{54}Cr$ abundance measured in meteorites and terrestrial planets.

3. Because of their small size, no direct mineralogical characterization of the $^{54}Cr$-rich grains could be performed. However, study by transmission electron microscopy shows that $^{54}Cr$-rich grains are found in a fraction where the sole chromium-bearing grains are oxides, mostly nanospinels (Fig. 2). Previous studies have shown that $^{54}Cr$ anomalies in meteorites can be released by leaching samples with hot hydrochloric acid. In our study, nanospinels less than 100 nm likely carrying $^{54}Cr$ excess were also digested by hot HCl treatment.

4. The large $^{54}Cr$-anomalies measured in meteorite residues can only be produced in supernovae (Fig. 12). At the present time, we cannot tell which of type Ia and II supernovae produced the grains identified here, though the oxide mineralogy would favor condensation from a type II. This question can be addressed in the future by measuring the isotopic abundance of other neutron-rich isotopes, in particular $^{48}Ca$, in the same grains. Study of these grains will shed new light on the nucleosynthesis of iron-group nuclei and the evolution of supernovae.

5.  It is possible that the same core-collapse supernova that synthesized $^{26}$Al and $^{60}$Fe present in the early solar system when meteorites formed also delivered the $^{54}$Cr-rich nanoparticles identified in this study. A test of this hypothesis will be to search for correlations between $^{54}$Cr isotopic anomalies and the abundances of extinct radioactivities of supernova origin (Fig. 13).

**Acknowledgements** Discussions with B.S. Meyer, F.K. Thielemann, R. Yokochi, A.M. Davis, and P.R. Craddock were appreciated. We thank T. Stephan and F.J. Stadermann for their help in our attempt to acquire Auger data on the $^{54}$Cr-rich grains. Constructive comments from an anonymous referee greatly improved the manuscript. The TEM national facility in Lille (France) is supported by the Conseil Regional du Nord-Pas de Calais, the European Regional Development Fund (ERDF), and the Institut National des Sciences de l'Univers (INSU, CNRS). J.H.C. and the laboratories at JPL were supported by NASA Cosmochemistry. This work was supported by a Packard fellowship, the France Chicago Center, a Moore Distinguished Scholarship at the California Institute of Technology, NASA and NSF through grants NNX09AG59G and EAR-0820807 to N.D.

**Figure captions**

**Fig. 1.** Chemical and physical methods used for concentrating the carrier of $^{54}$Cr anomalies (see Section 2 for details). The fractions enclosed in boxes were studied by TEM, TIMS and NanoSIMS. *O* and *M* stand for Orgueil and Murchison, respectively. *Colloid*, *>200*, *200-800*, and *<800* are colloidal, <200 nm, 200-800 nm, and >800 nm nominal size fractions, respectively.

**Fig. 2.** TEM micrographs. **A.** Amorphous, fluffy organic matrix in the colloidal fraction of Orgueil containing small diffracting particles (dark dots) of oxide grains (generally spinels). **B, C.** Small spinel grains partially embedded in the organic matrix of the colloidal fraction of Murchison. **D.** Amorphous carbon nanoglobule surrounded by fluffy organic matter containing spinel grains (one is in Bragg condition). This nanoglobule was found in the colloidal fraction of Murchison but similar globules were found in Orgueil. **E, F.** Large spinel grains found in the >200 nm fraction of Orgueil.

**Fig. 3.** X-ray elemental maps of a large clump of organic matter and mineral grains found in the colloidal fraction of Murchison. Exactly the same features are found in the colloidal fraction of Orgueil. Chromium is distributed in small grains containing various amounts of Al, Mg, and Fe. Sulfur and phosphorus (not shown here) are homogeneously distributed in the clump.

**Fig. 4.** Spinel composition as a function of grain size (Table 1). Normalized atomic fractions of aluminum and chromium are presented (normalized to 100 % non-oxygen atoms in the spinel crystal) for the colloidal and <200 nm size fractions of both Orgueil and Murchison residues. Small spinels are richer in Al and poorer in Cr than large spinels.

**Fig. 5.** TEM bright field micrographs of the colloidal fraction from Orgueil before (**A**) and after (**B**) leaching with 5 mol/L HCl for a day at 80 °C. These two micrographs are representative of the whole samples. Diffracting crystals embedded in the organic matrix appear as dark particles and are circled. For clarity, only the largest particles are circled in panel A. The grains are mainly Cr-bearing spinels. Comparison of A and B shows that nano-spinels can be dissolved in hot HCl, explaining why $^{54}$Cr-excess has been found previously in those leachates (Rotaru et al. 1992; Podosek et al. 1997; Nichols et al. 1998; 2000; Alexander 2002; Shukolyukov & Lugmair 2006; Trinquier et al. 2007; Birck et al. 2009; Qin et al. 2010).

**Fig. 6.** Quality assessment of $^{54}$Cr/$^{52}$Cr isotope ratio images. Comparison between images generated by *L'image* software of L.R. Nittler (**A**), images generated with a *Mathematica* script written by the authors (**B**) and synthetic noise images calculated using measured $^{52}$Cr$^+$ and $^{56}$Fe$^+$ and assuming constant $^{54}$Cr/$^{52}$Cr and $^{54}$Fe/$^{56}$Fe ratios (**C**). In the synthetic images, the total counts at each pixel were randomized using Poisson distribution and the synthetic data were processed using the same algorithm as that applied to real data. By construction, the synthetic image should show no anomaly and all isotopic variations can be ascribed to counting statistics. In all cases, ion counts were averaged over moving boxes of 11 pixel width. The masks (black in *L'image* and white in *Mathematica*) correspond to regions where the $^{54}$Fe correction on $^{54}$Cr exceeds 20,000 ε or where the $^{52}$Cr$^+$ ion counts represent less than 1 % of the maximum ion count in the image. The top and bottom images are 40×40 and 10×10 μm$^2$, respectively.

**Fig. 7.** Chromium isotopic compositions of bulk residues from (**A**) Orgueil and (**B**) Murchison (Table 3). The yellow diamonds are the measured ε$^{54}$Cr values ($^{54}$Cr/$^{52}$Cr ratio normalized to solar on the right axis) plotted against the median size of the grains. The blue curves are the size distributions of Cr-bearing grains in each nominal size fraction. The $^{54}$Cr/$^{52}$Cr ratio was measured by TIMS (mass fractionation was corrected by internal normalization using the $^{50}$Cr/$^{52}$Cr ratios). The size-distributions of Cr-bearing grains in the colloidal and <200 nm nominal size fractions were measured by TEM. The size distributions in the 200-800 nm and >800 nm size fractions were measured by SEM. The fact that the isotopic

anomalies are larger in the colloidal fraction (median size ~30 nm) compared to coarser fractions indicates that the carrier of $^{54}$Cr anomalies must be very fine (<100 nm).

**Fig. 8.** $\varepsilon^{54}$Cr data of ROIs from Orgueil and Murchison residues. Only ROIs with Fe/Cr<2 are shown (corresponding to a maximum $^{54}$Fe correction on $^{54}$Cr of ~17,000 $\varepsilon$). Mass fractionation was corrected by internal normalization to a constant $^{53}$Cr/$^{52}$Cr ratio of 0.11339 using the exponential law. Isotopic ratios of individual ROIs are normalized to the composition of the bulk image, where Fe-rich regions have been excluded. The two $^{54}$Cr-rich spots identified in this study are not shown.

**Fig. 9.** 40×40 µm$^2$ ion and isotopic ratio images of $^{54}$Cr-rich grain np1 (delineated by a thin white line) and its surroundings. See Fig. 6 caption for details on the image processing.

**Fig. 10.** 10×10 µm$^2$ ion and isotopic ratio images of $^{54}$Cr-rich grain np1 (delineated by a thin white line) and its surroundings. See Fig. 6 caption for details on the image processing.

**Fig. 11.** $^{54}$Cr-rich spot in a residue (<200 nm nominal size fraction) from the Orgueil meteorite (nanoparticle np1, Table S2). The red spot in the larger image has $^{54}$Cr/$^{52}$Cr =3.6×solar ($\varepsilon^{54}$Cr=+25,861). The inset shows another image acquired on a different day at the same location that confirms the presence of a $^{54}$Cr-rich grain ($^{54}$Cr/$^{52}$Cr =2.8×solar, $\varepsilon^{54}$Cr=+17,902). The lower $^{54}$Cr/$^{52}$Cr ratio measured the second time is most likely due to the fact that the grain was partially consumed during the first analysis. Ion counts were averaged over a moving box of 11 pixel width. The white mask correspond to regions where the $^{54}$Fe correction on $^{54}$Cr exceeds 20,000 $\varepsilon$ or where the $^{52}$Cr$^+$ ion counts represent less than 1 % of the maximum ion count in the image.

**Fig. 12.** Comparison between the Cr isotopic composition measured in the $^{54}$Cr-rich grain np1 (Fig. 2) and model predictions for type Ia and II supernovae. Both types of supernovae can explain the $^{54}$Cr/$^{52}$Cr ratio measured in $^{54}$Cr-rich grain np1 (~3.6×solar, which represents a lower limit on the actual ratio due to dilution with solar Cr during the measurement). **A.** The SNIa predictions are for a low flame speed and a range of central ignition densities between 2×10$^9$ g/cm$^3$ and 8×10$^9$ g/cm$^3$ (Woosley 1997). The predicted $^{54}$Cr/$^{52}$Cr ratios are probably too high because the electron capture rates used by Woosley (1997) have been subsequently revised downward (Brachwitz et al. 2000). No comprehensive study of the nucleosynthesis of neutron-rich isotopes in SNIa has been published with these reduced rates but a calculation with a central ignition density of 6×10$^9$ g/cm$^3$ gives a $^{54}$Cr/$^{52}$Cr ratio

of ~9×solar (Thielemann et al. 2004). **B.** SNII predictions are for a 21 $M_\odot$ progenitor star (Rauscher et al. 2002; http://nucastro.org/nucleosynthesis/). Only the zones where neutron-capture was active (O/C and O/Ne, between 2.5 and 5.3 $M_\odot$ in mass coordinate) are shown.

**Fig. 13.** Predicted correlation between $\varepsilon^{54}$Cr and the initial $^{60}$Fe/$^{56}$Fe ratio if $^{54}$Cr-rich nanoparticles were produced in the same core-collapse supernova that injected $^{60}$Fe in the early solar system (see text for details). The prediction is that for the range of $\varepsilon^{54}$Cr values measured in bulk meteorites (from -0.8 in achondrites to +1.6 in CI chondrites; Trinquier et al. 2007; Qin et al. 2010), variations in the initial $^{60}$Fe/$^{56}$Fe ratio should be on the order of 10-20 %. The results obtained so far provide no evidence for $^{60}$Fe heterogeneity at the scale of bulk planetary objects outside of ±10 % (Dauphas et al. 2008).

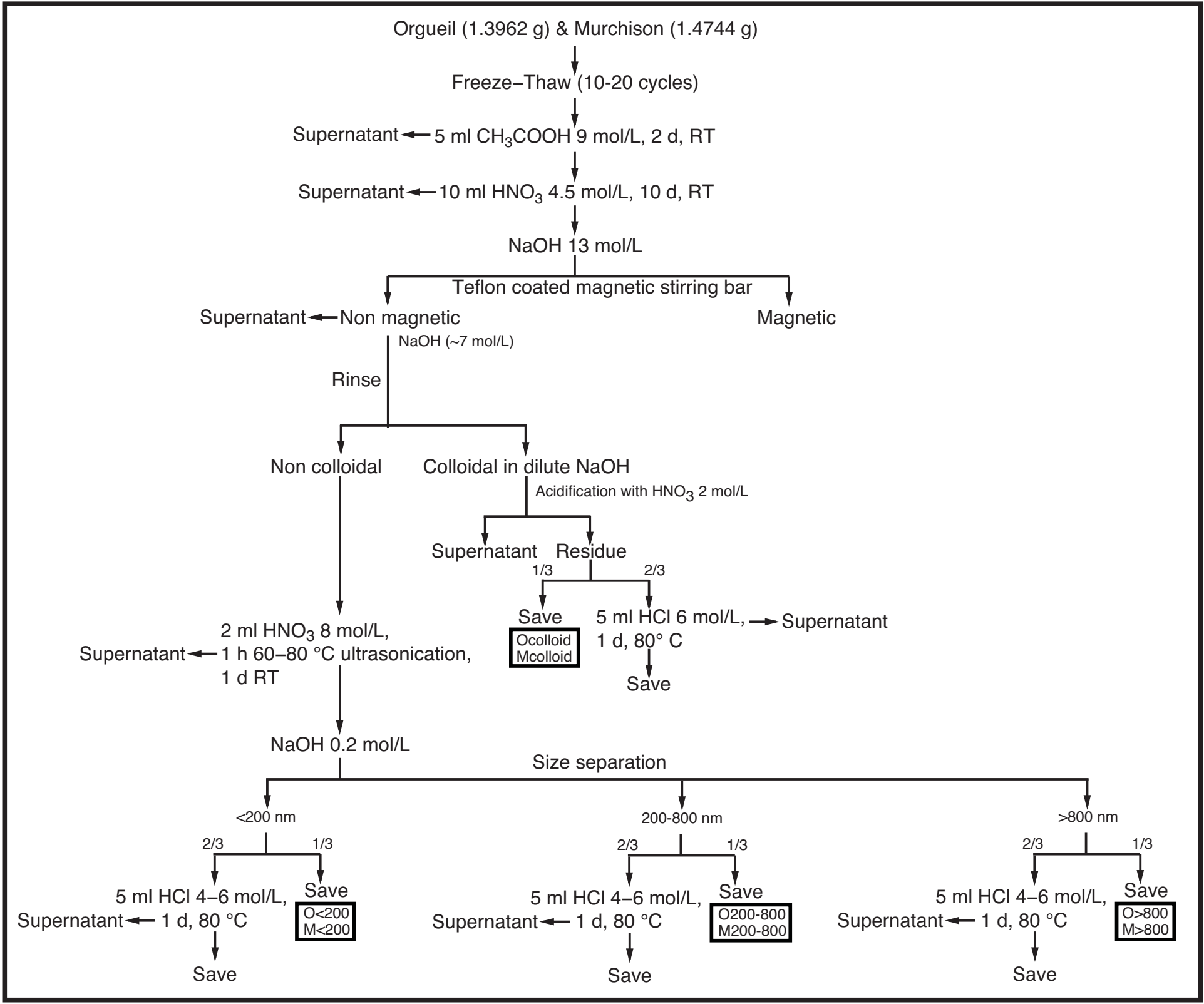

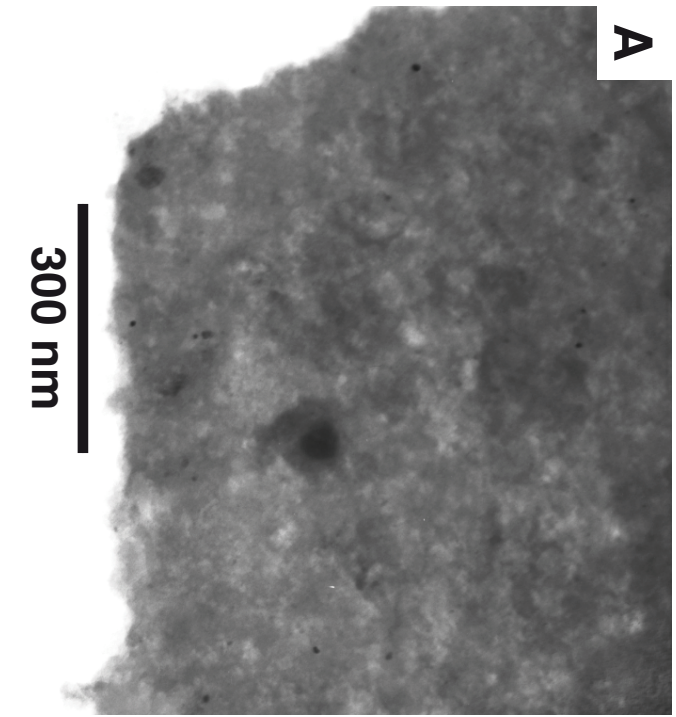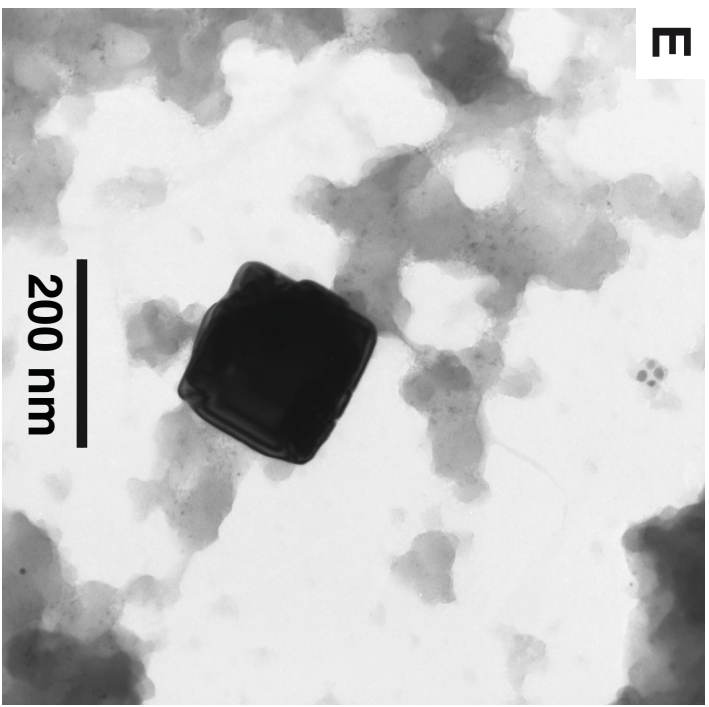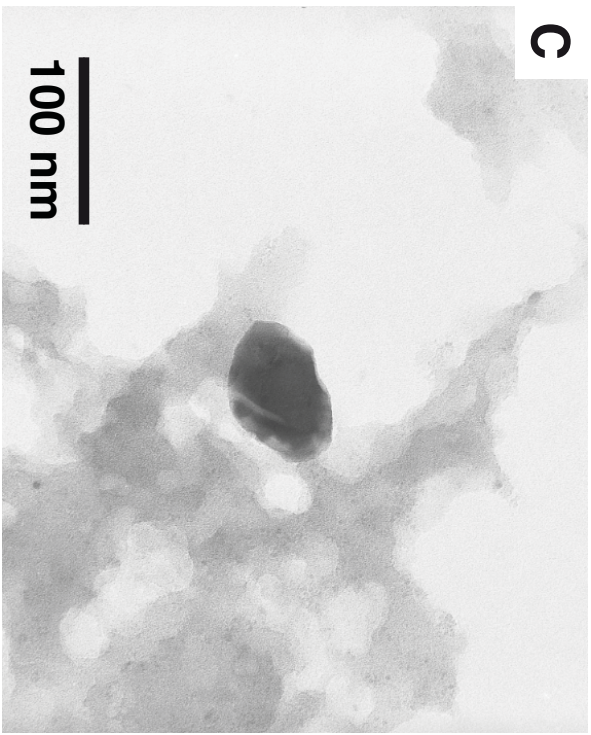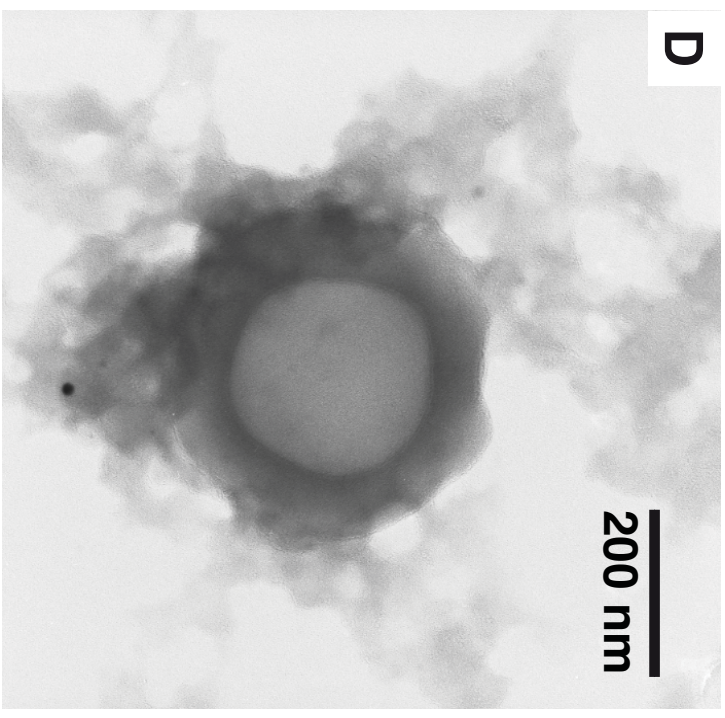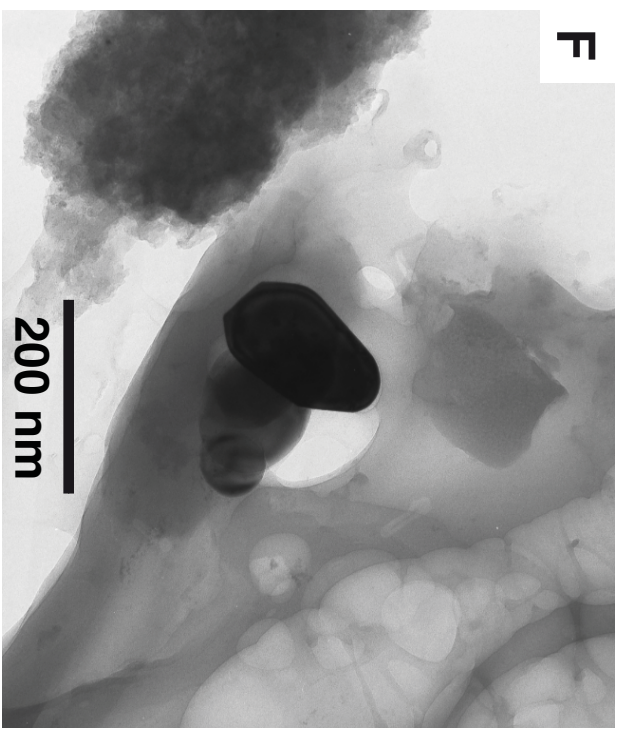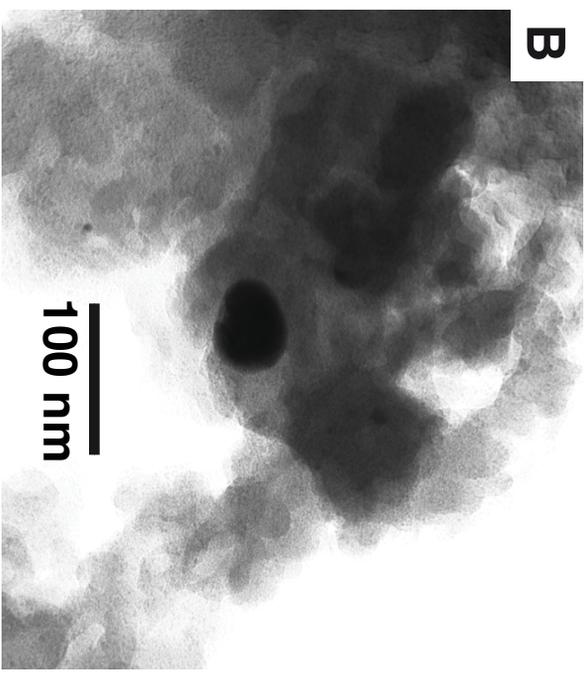

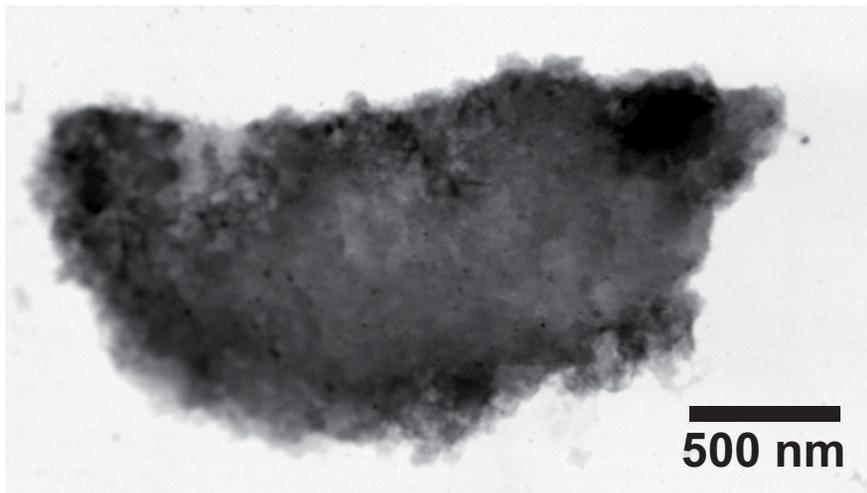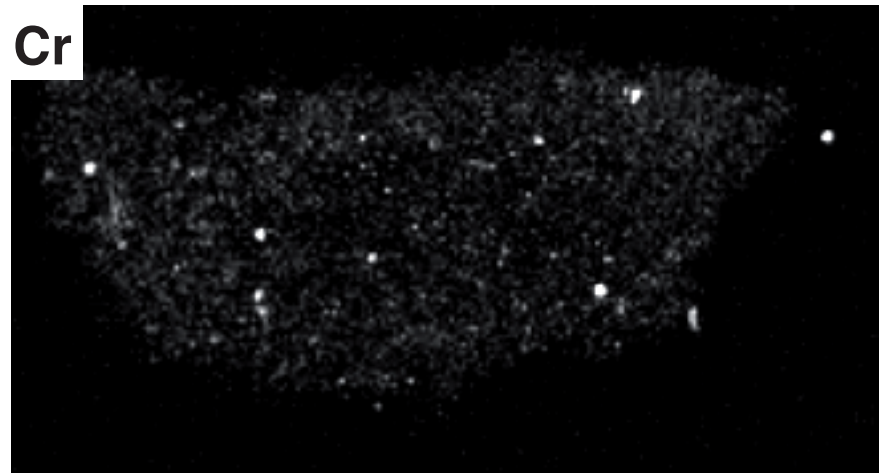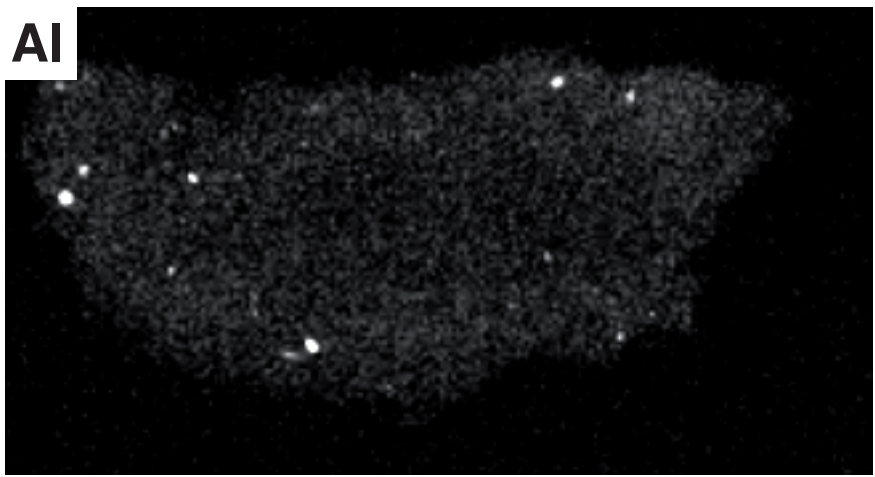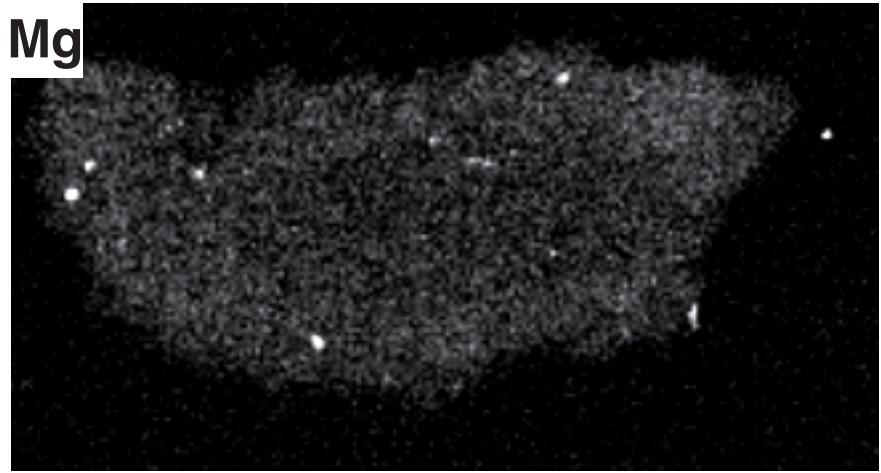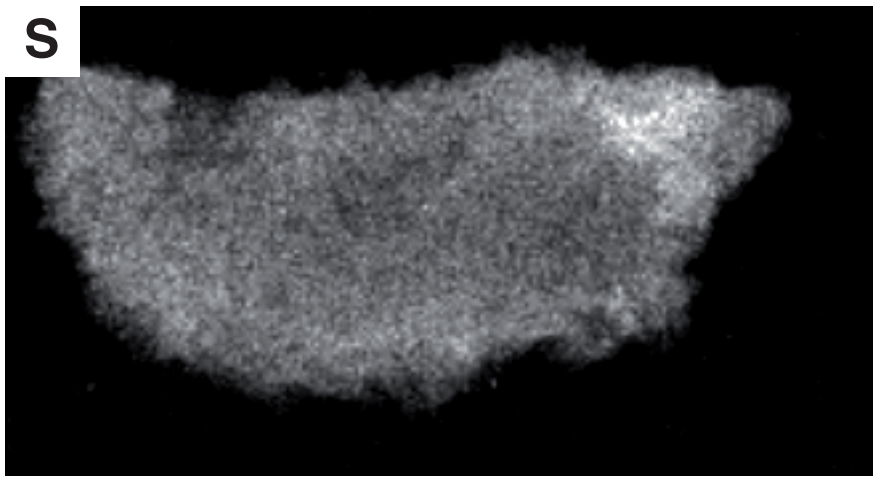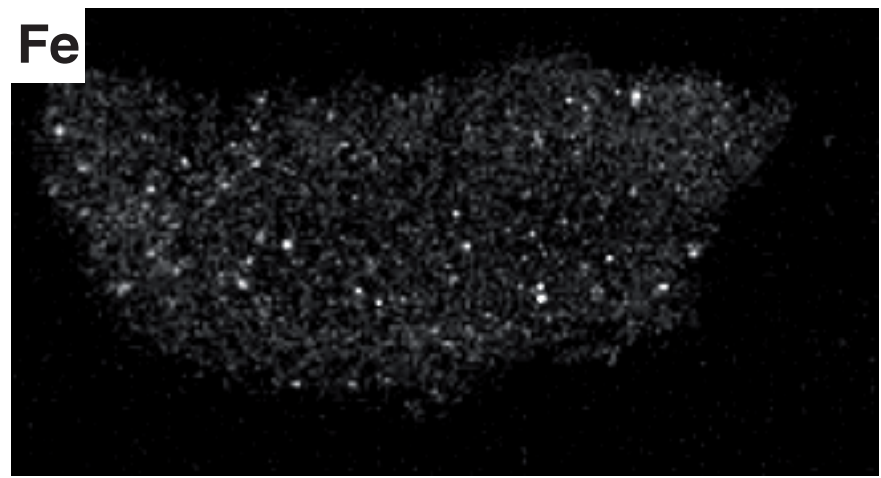

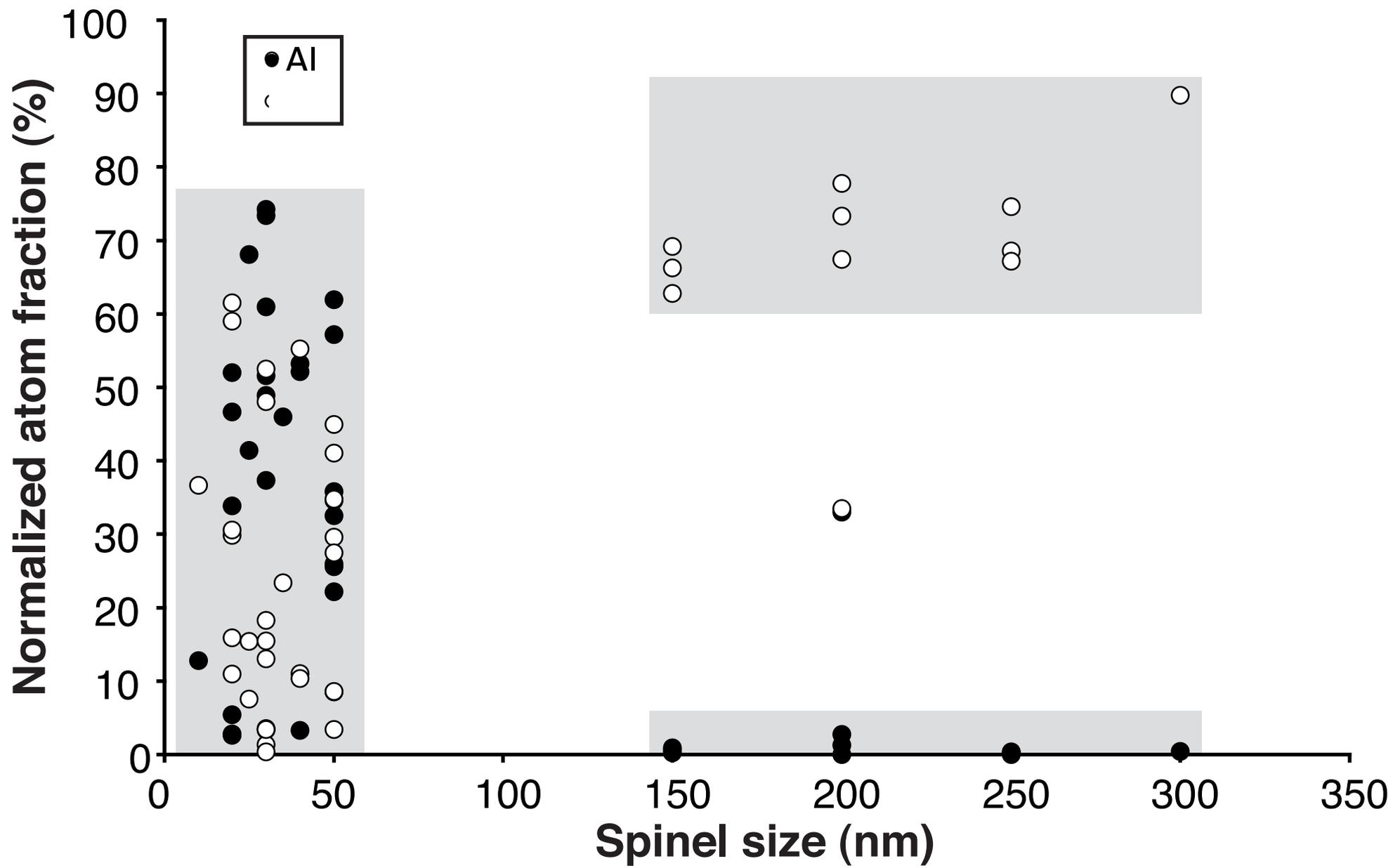

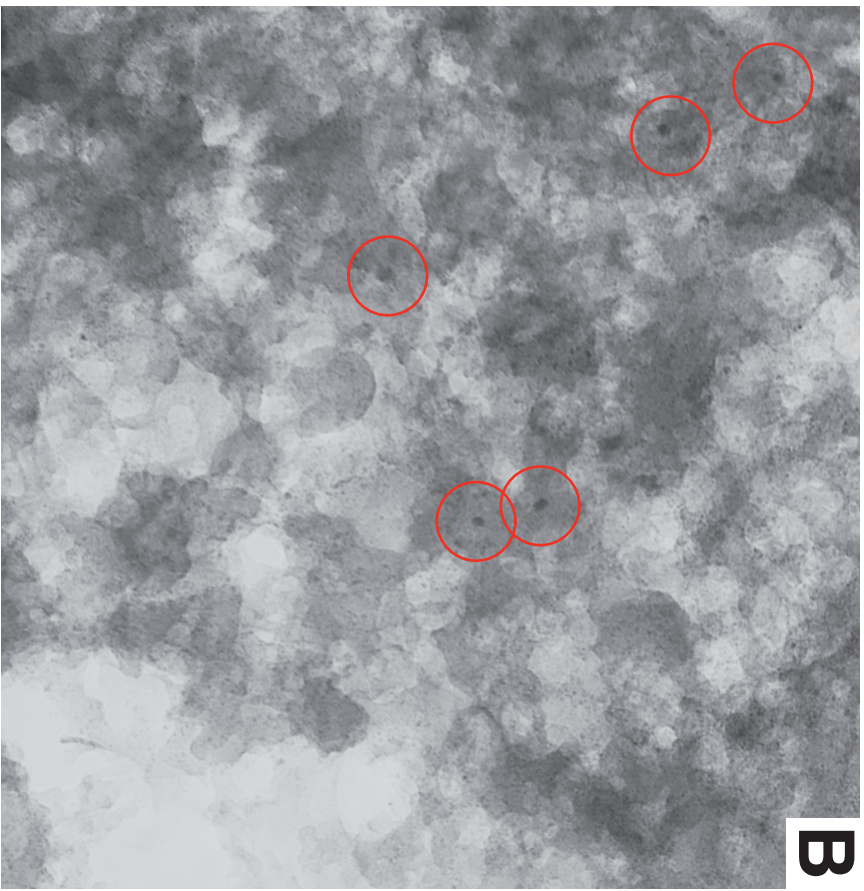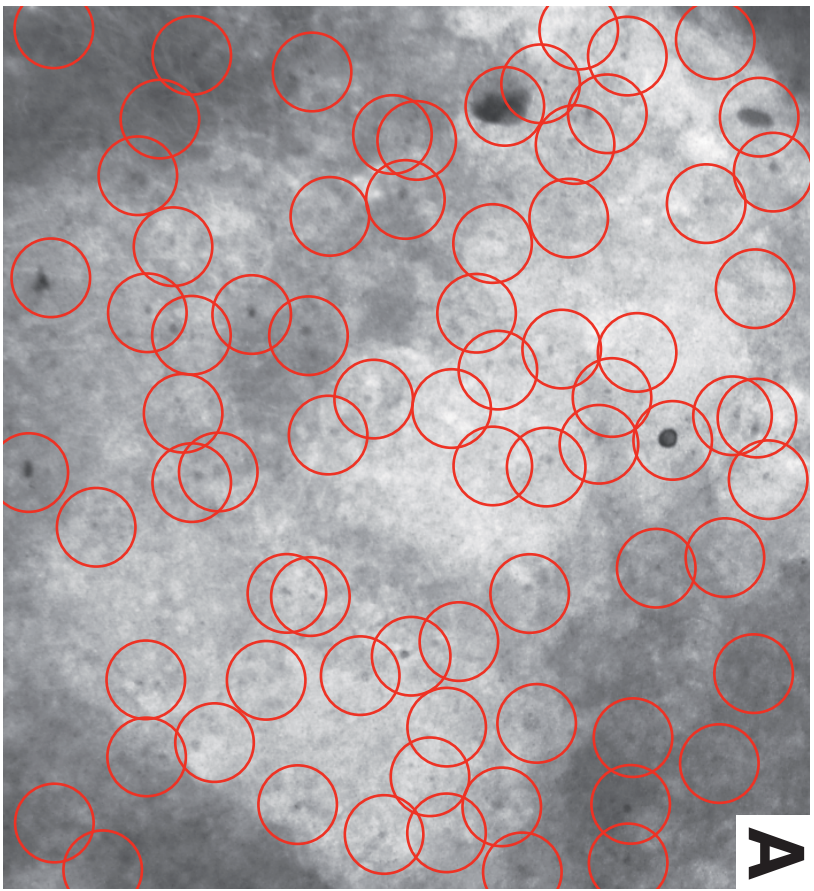

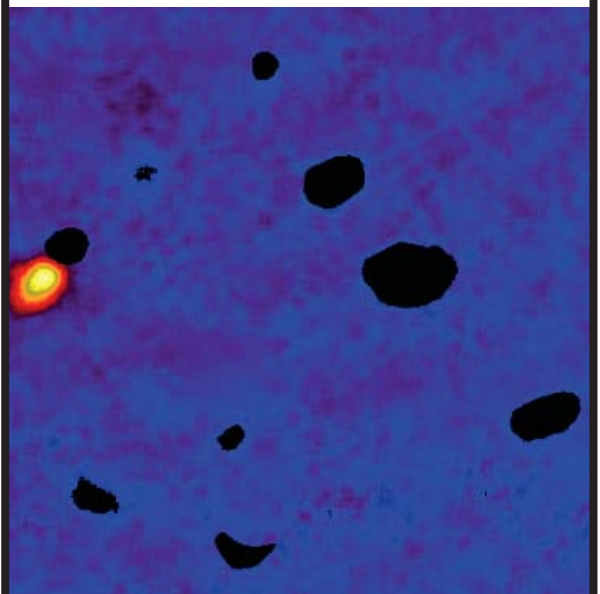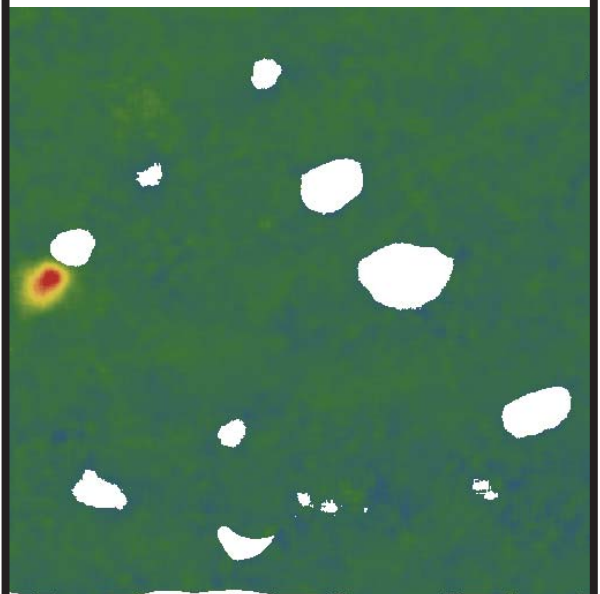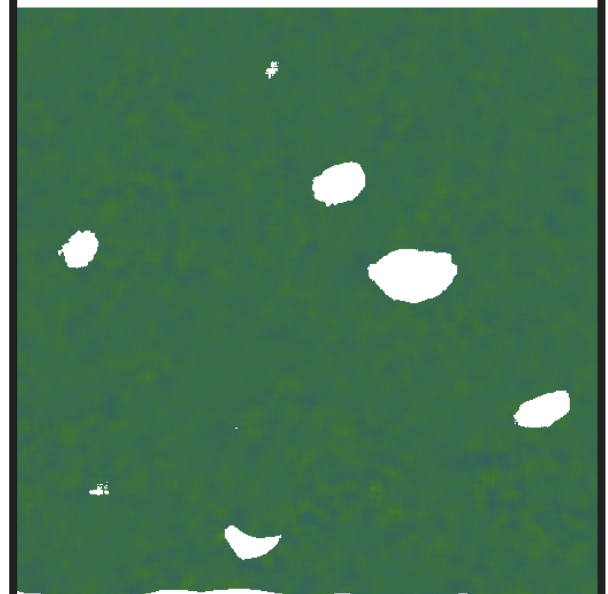

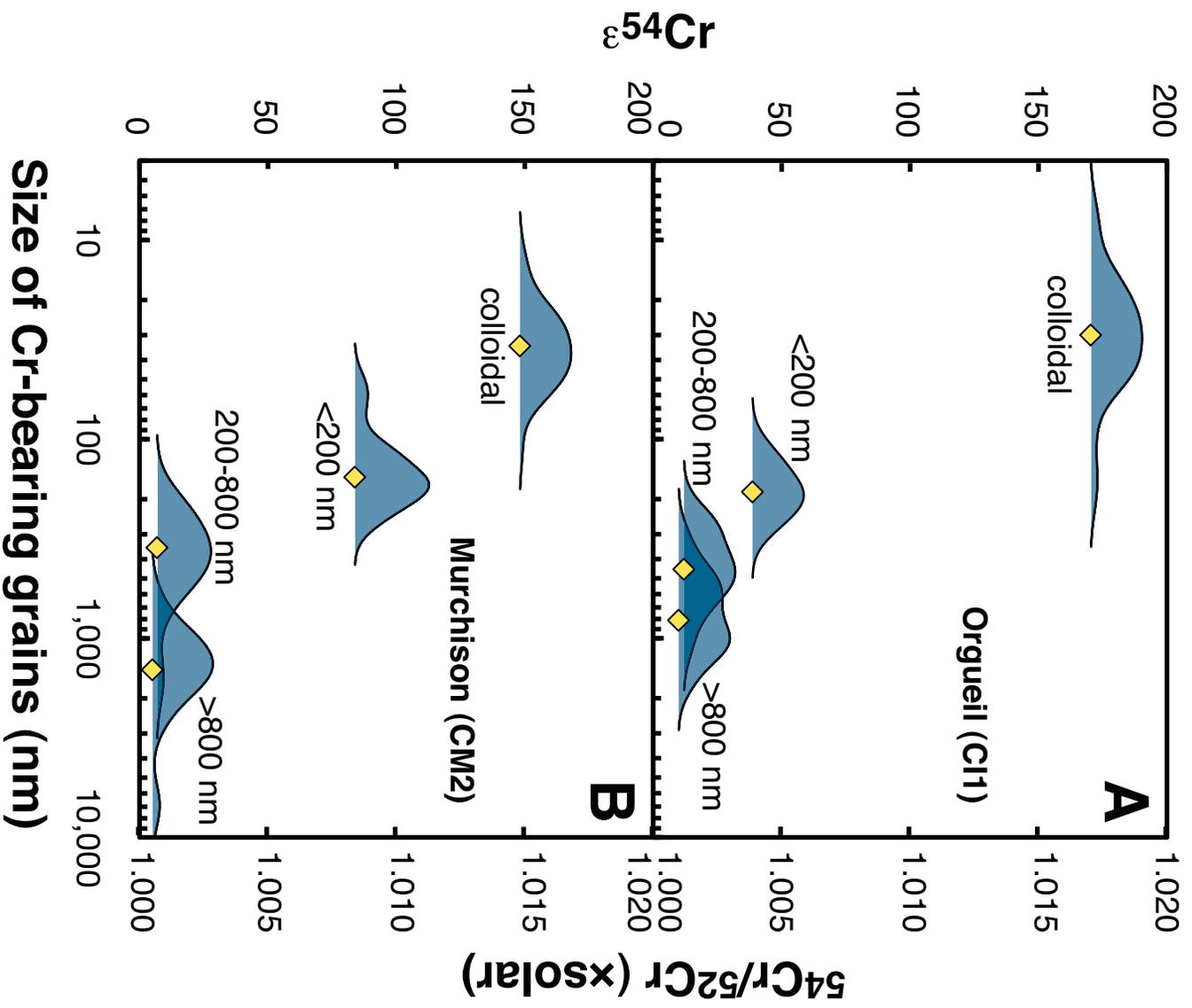

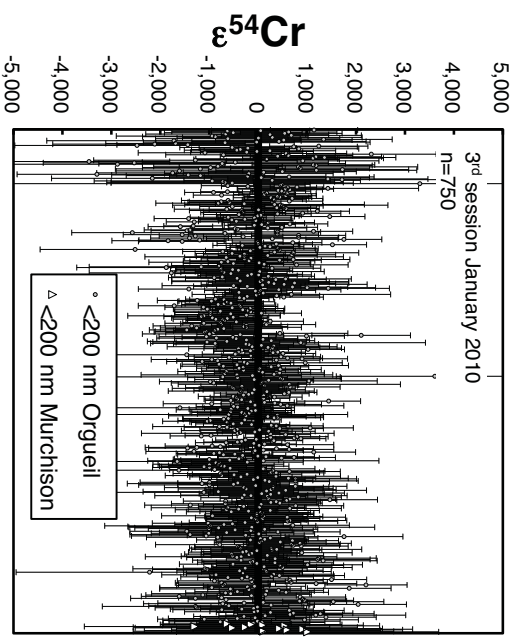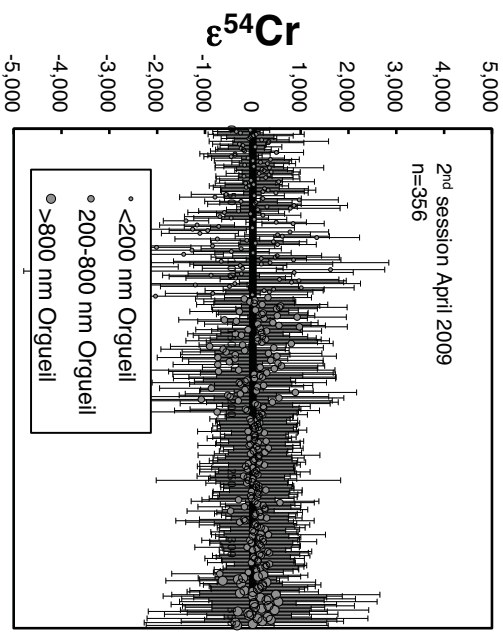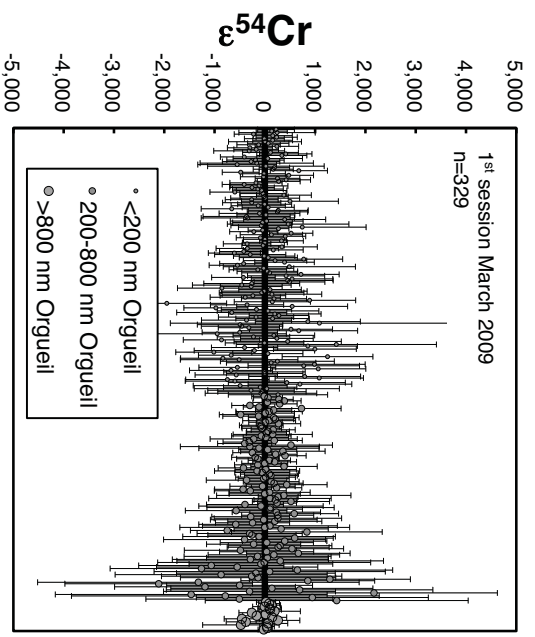

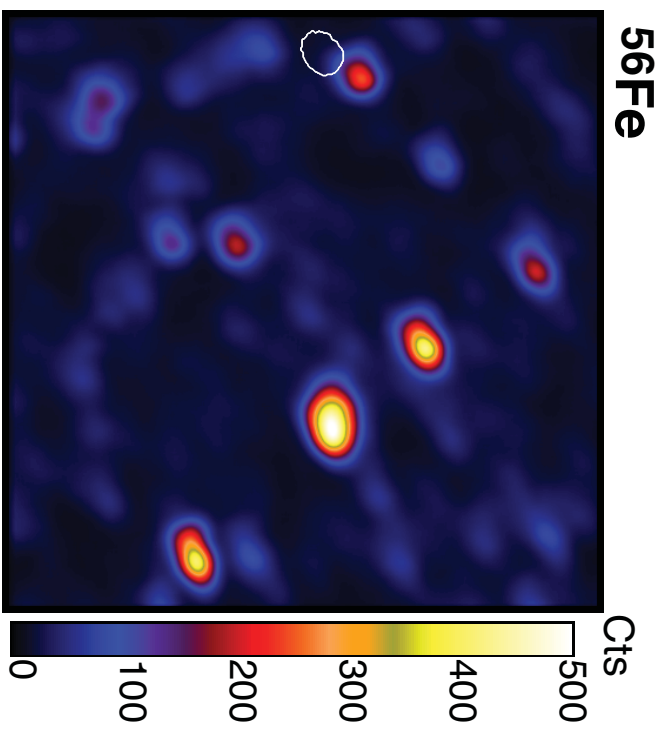 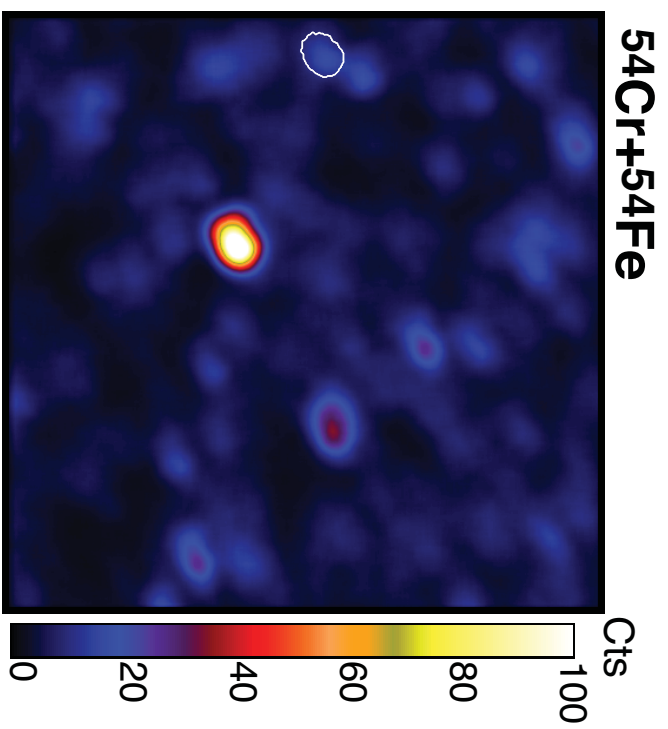 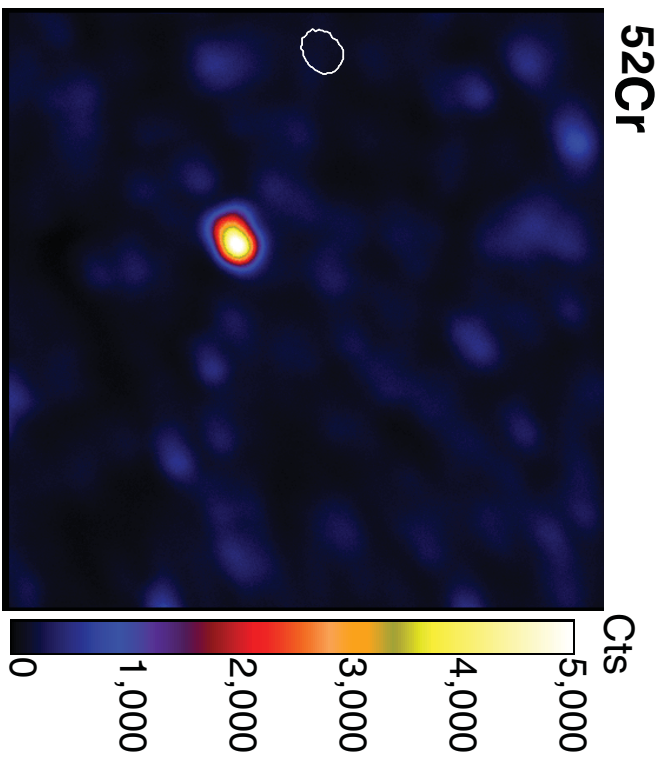
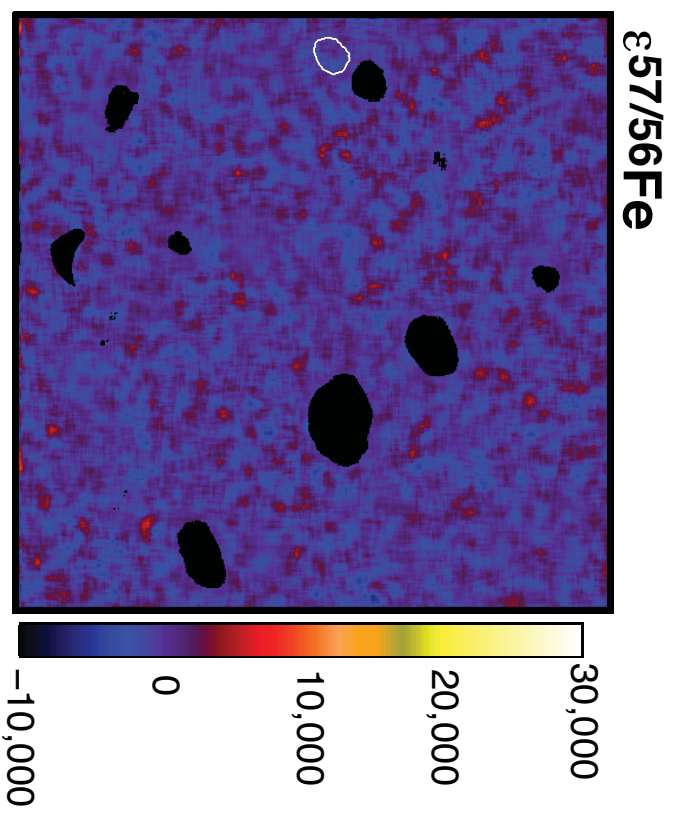 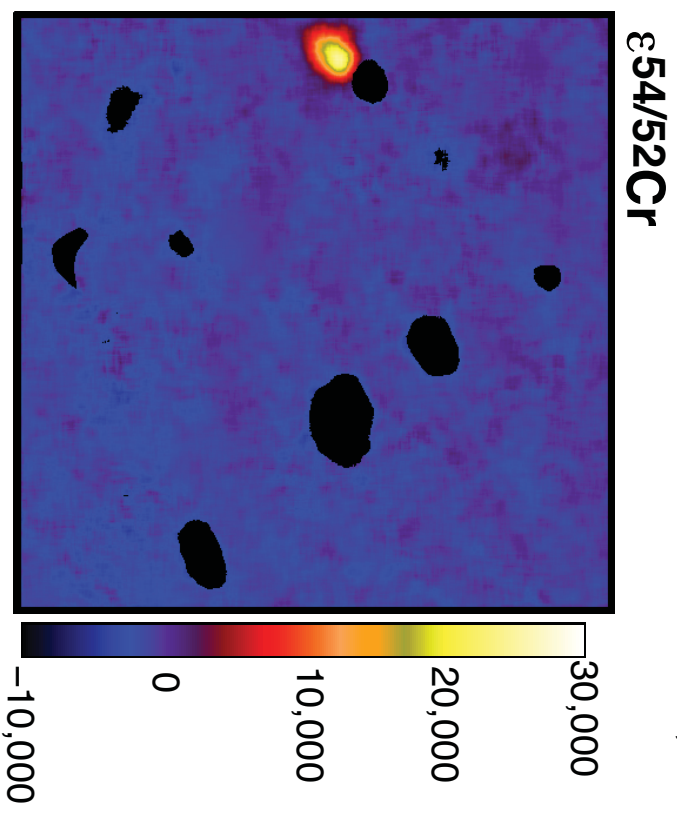 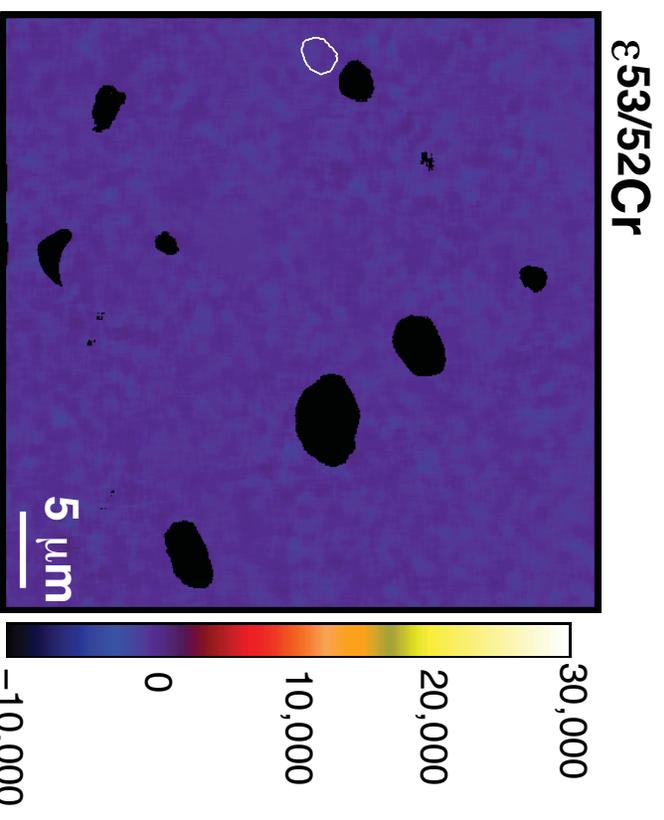

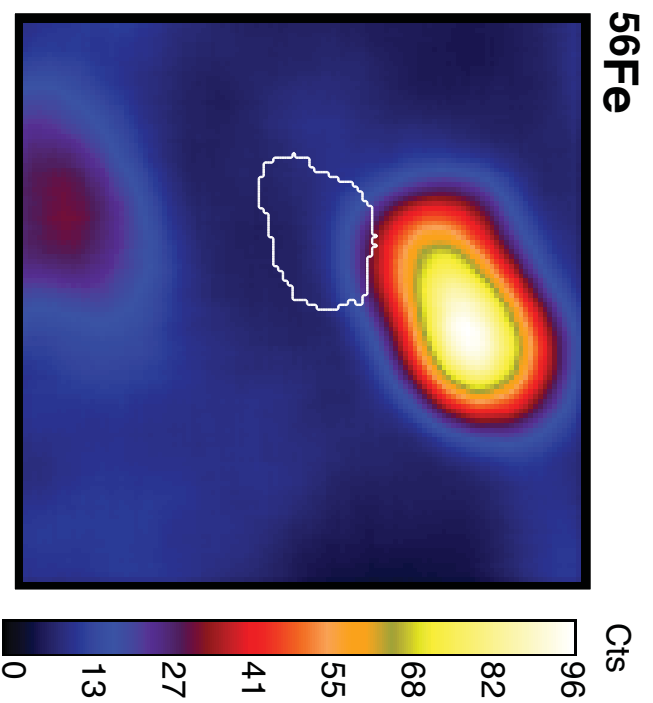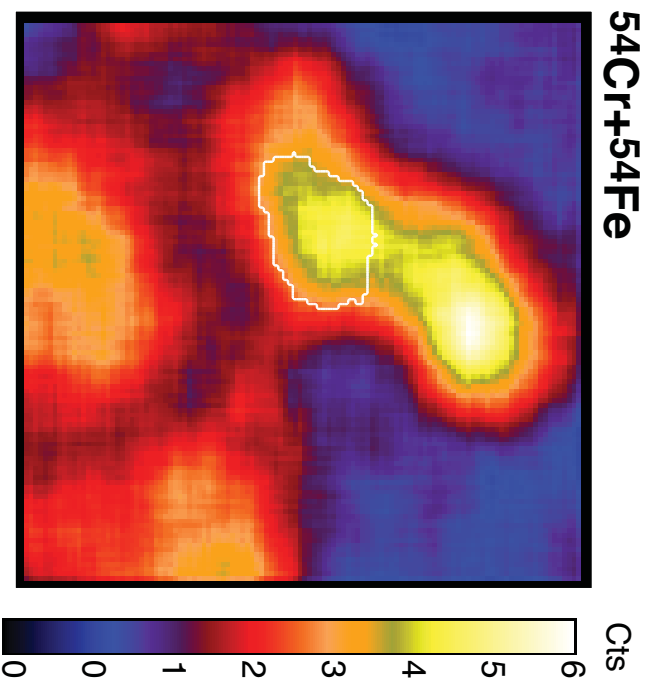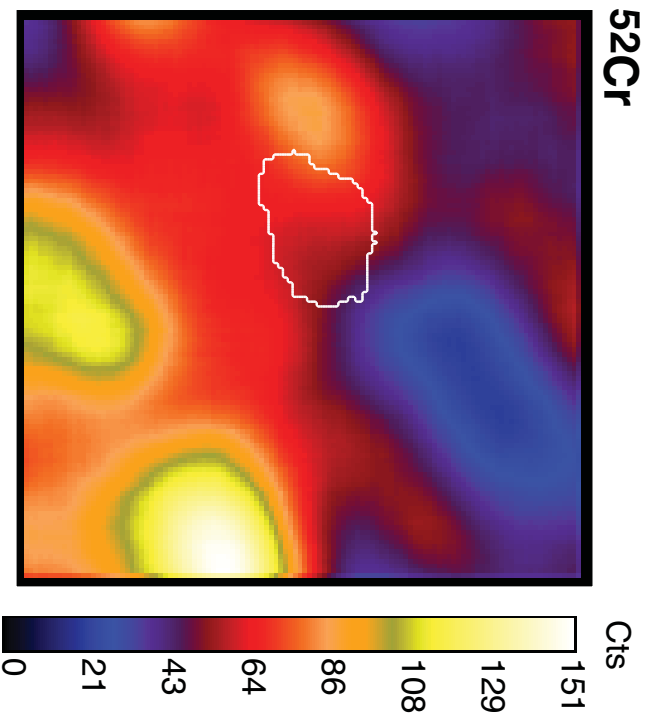
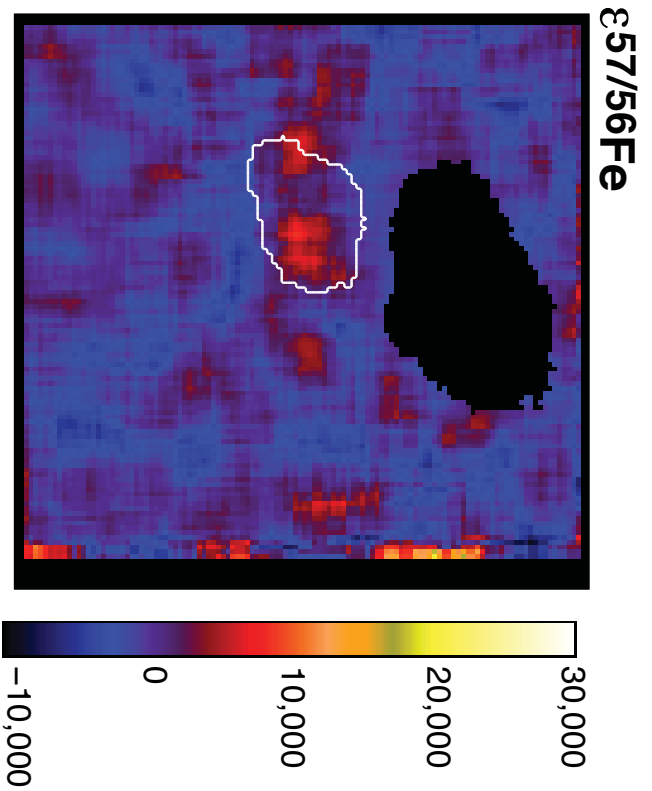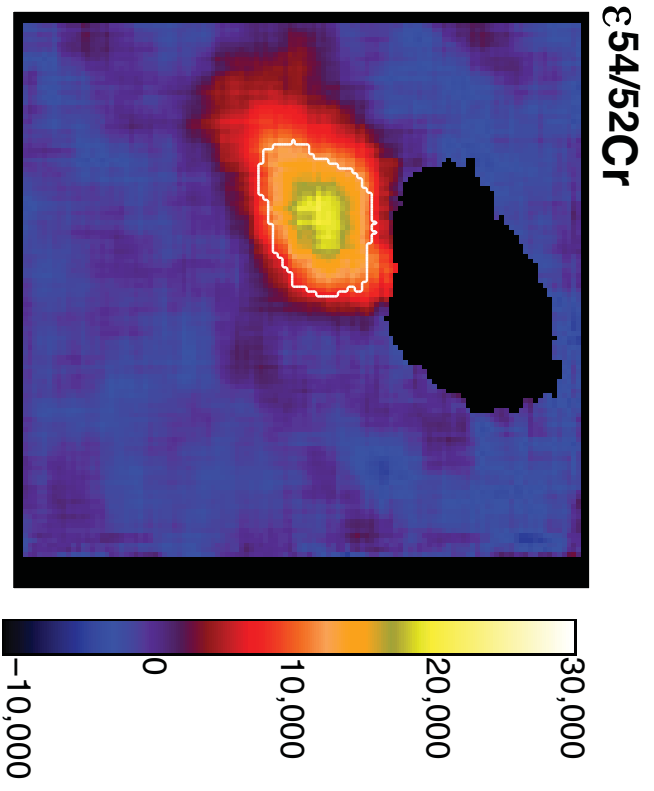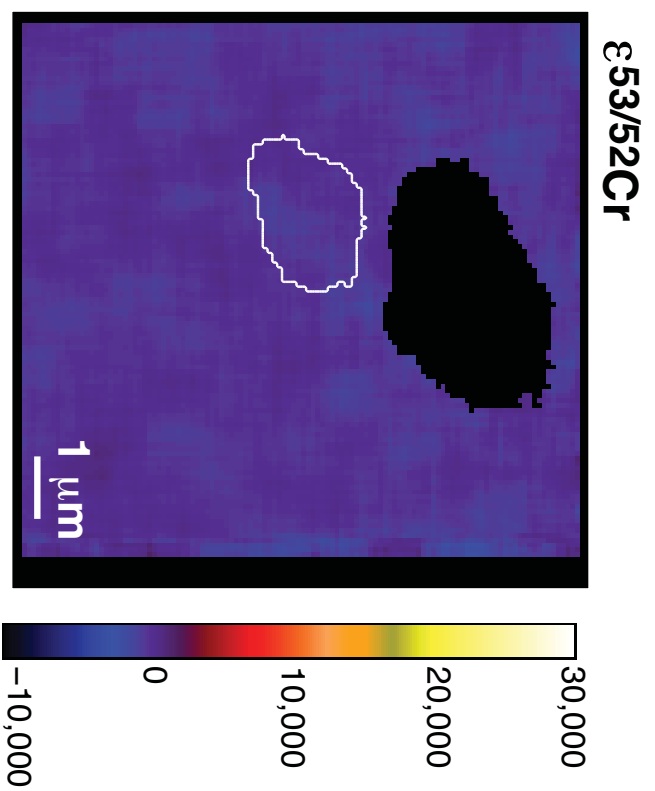

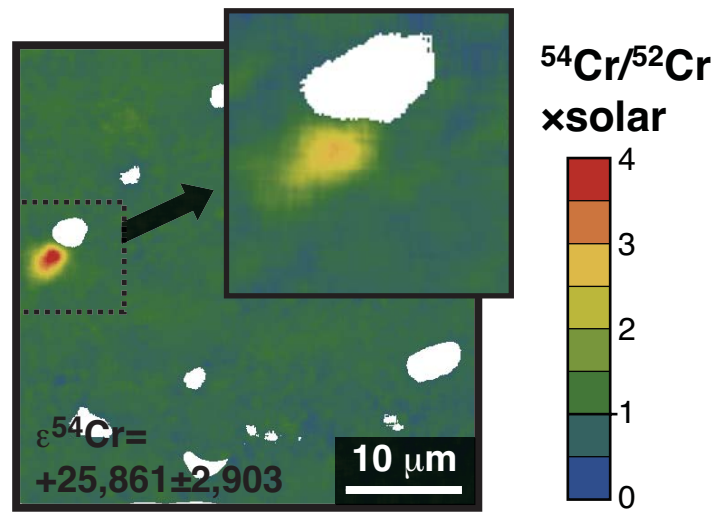

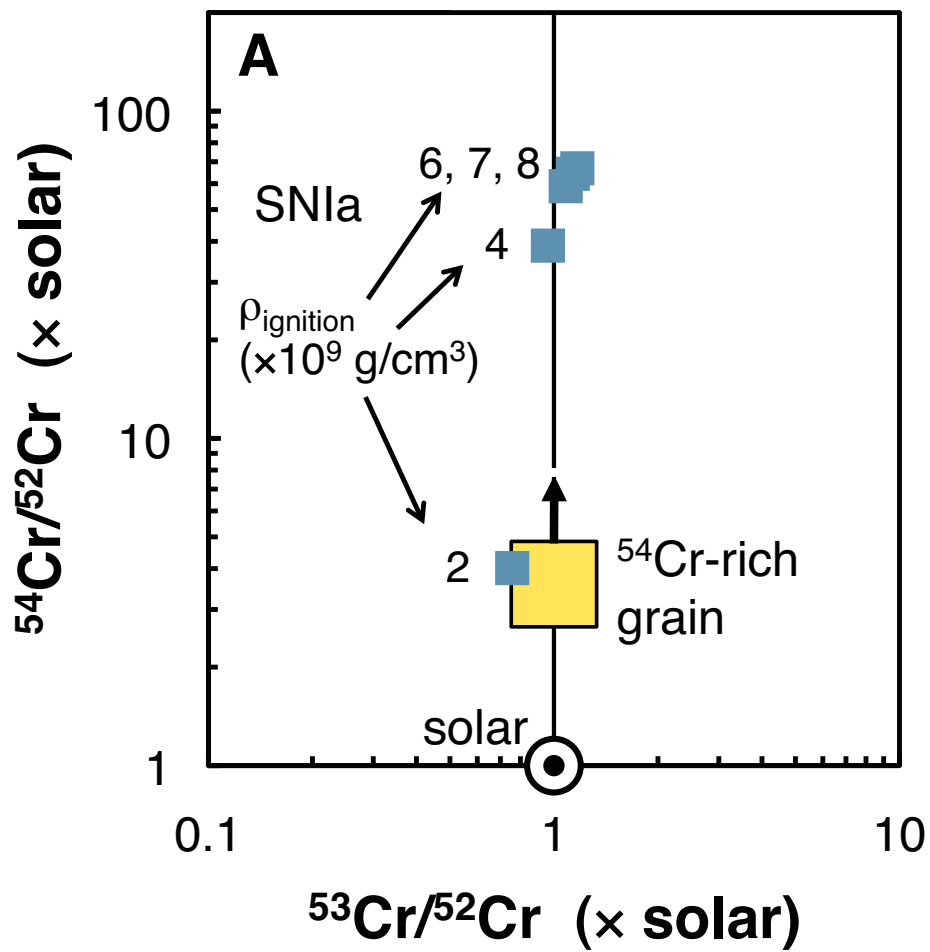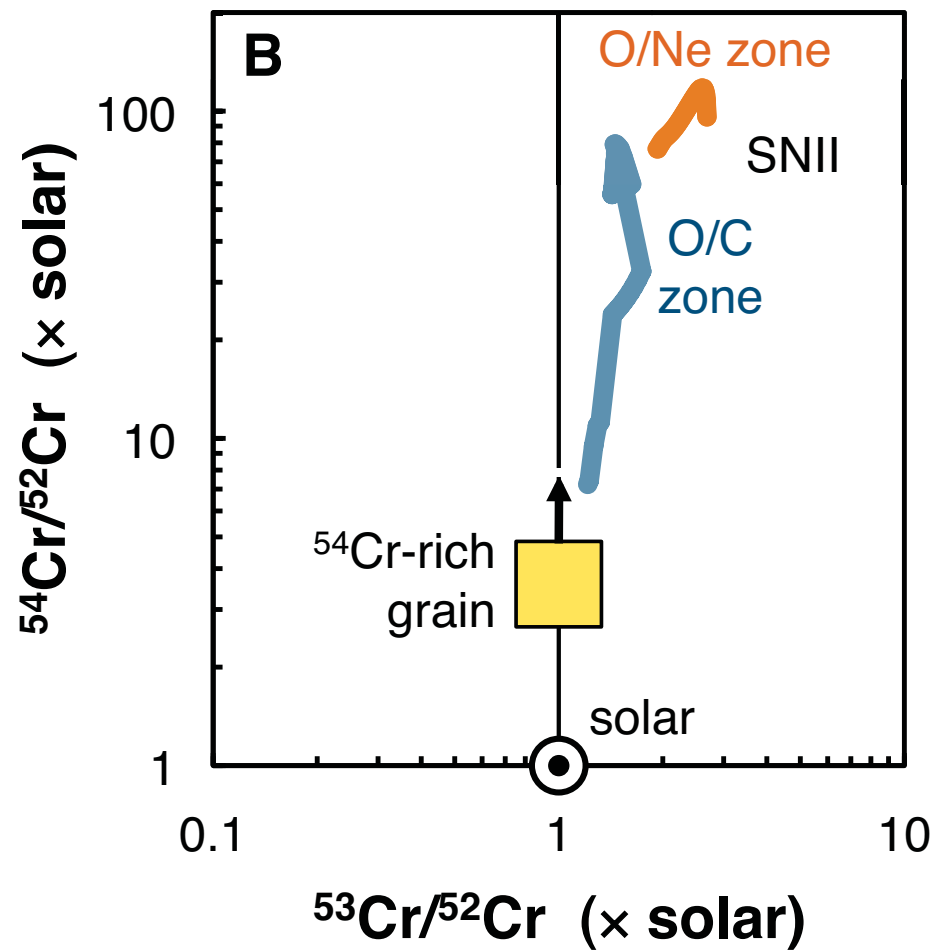

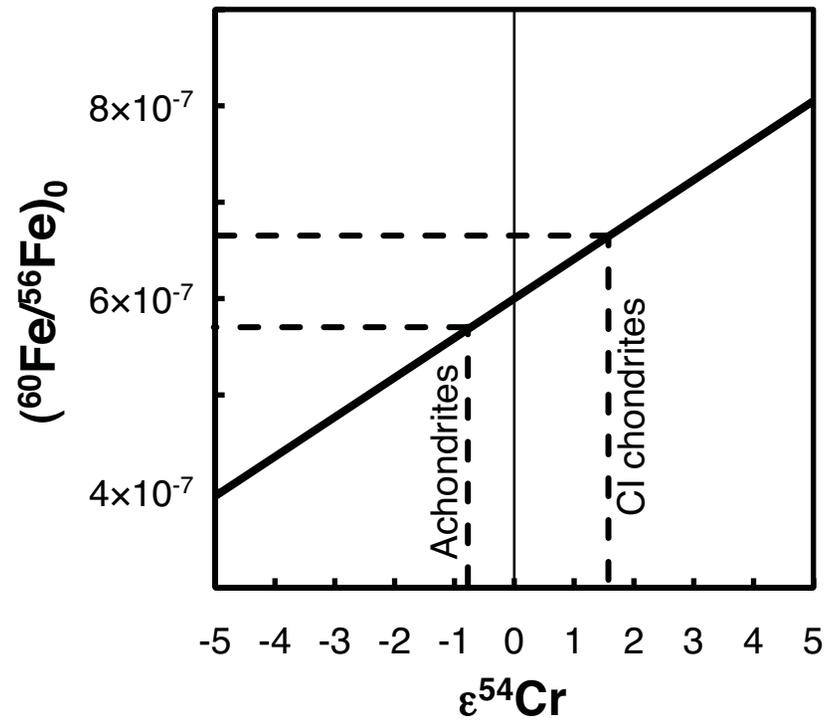

**Table 1**
Atomic compositions of nanospinels (normalized to 100 % non-oxygen atoms)

| Meteorite | Size fraction | size (nm) | Cr | Al | Mg | Fe |
|---|---|---|---|---|---|---|
| Orgueil (CI1) | | | | | | |
| | Colloidal | 10 | 36.63 | 12.77 | 27.26 | 23.33 |
| | Colloidal | 20 | 58.99 | 2.78 | 35.04 | 3.19 |
| | Colloidal | 20 | 29.82 | 33.88 | 10.25 | 26.04 |
| | Colloidal | 20 | 15.91 | 46.64 | 36.24 | 1.21 |
| | Colloidal | 30 | 1.33 | 73.37 | 25.06 | 0.23 |
| | Colloidal | 30 | 3.43 | 74.26 | 15.69 | 6.62 |
| | Colloidal | 30 | 48.07 | 3.56 | 27.98 | 20.39 |
| | Colloidal | 35 | 23.36 | 45.99 | 21.61 | 9.05 |
| | Colloidal | 40 | 55.27 | 3.31 | 27.59 | 13.84 |
| | Colloidal | 40 | 10.35 | 53.21 | 35.47 | 0.96 |
| | Colloidal | 50 | 34.59 | 22.14 | 32.66 | 10.61 |
| | Colloidal | 50 | 34.75 | 32.54 | 32.08 | 0.63 |
| | Colloidal | 50 | 27.46 | 35.82 | 32.85 | 3.87 |
| | Colloidal | 50 | 3.42 | 61.96 | 34.62 | 0.00 |
| | Colloidal | 50 | 41.03 | 25.59 | 10.76 | 22.63 |
| | Colloidal | 150 | 66.28 | 0.91 | 19.93 | 12.88 |
| | <200 nm | 150 | 69.21 | 0.21 | 19.66 | 10.91 |
| | <200 nm | 200 | 77.77 | 2.72 | 3.66 | 15.85 |
| | <200 nm | 200 | 67.41 | 1.27 | 14.14 | 17.18 |
| | <200 nm | 250 | 67.20 | 0.35 | 28.72 | 3.73 |
| | <200 nm | 250 | 74.60 | 0.29 | 0.32 | 24.79 |
| Murchison (CM2) | | | | | | |
| | colloidal | 20 | 30.55 | 5.45 | 49.16 | 14.84 |
| | colloidal | 20 | 10.94 | 51.98 | 33.34 | 3.73 |
| | colloidal | 20 | 61.48 | 2.62 | 12.24 | 23.65 |
| | colloidal | 25 | 7.58 | 68.09 | 15.18 | 9.14 |
| | colloidal | 25 | 15.40 | 41.40 | 12.86 | 30.34 |
| | colloidal | 30 | 13.04 | 51.60 | 25.26 | 10.10 |
| | colloidal | 30 | 0.38 | 60.94 | 37.33 | 1.35 |
| | colloidal | 30 | 18.27 | 37.33 | 13.49 | 30.92 |
| | colloidal | 30 | 15.48 | 48.90 | 10.47 | 25.15 |
| | colloidal | 30 | 52.52 | 3.35 | 32.29 | 11.84 |
| | colloidal | 40 | 11.05 | 52.12 | 36.48 | 0.35 |
| | colloidal | 50 | 44.94 | 8.53 | 27.41 | 19.12 |
| | colloidal | 50 | 29.61 | 25.94 | 31.52 | 12.93 |
| | colloidal | 50 | 8.57 | 57.19 | 34.07 | 0.17 |
| | colloidal | 150 | 62.80 | 0.58 | 30.10 | 6.52 |
| | <200 nm | 200 | 33.52 | 33.00 | 32.39 | 1.10 |
| | <200 nm | 200 | 73.34 | 0.00 | 8.74 | 17.92 |
| | <200 nm | 250 | 68.56 | 0.00 | 26.30 | 5.14 |
| | <200 nm | 300 | 89.80 | 0.43 | 0.43 | 9.34 |

**Table 2**
Cup configurations for Cr isotope measurements by TIMS

| Configuration | Acquisition time | Faraday cups | | | | |
| --- | --- | --- | --- | --- | --- | --- |
| | | L4 | L2 | C | H2 | H4 |
| Line 1 | 33 s | | $^{52}$Cr | $^{53}$Cr | $^{54}$Cr | $^{56}$Fe |
| Line 2 | 33 s | $^{48}$Ti | $^{50}$Cr | $^{51}$V | $^{52}$Cr | |

**Table 3**

Chromium isotopic compositions of meteorite residues

| | | Median size of Cr-bearing grains (nm) | $^{55}$Mn/$^{52}$Cr | $\varepsilon^{53}$Cr | $\varepsilon^{54}$Cr |
|---|---|---|---|---|---|
| Orgueil residues | Colloidal | 30 | 0.07 | -2.17 ± 0.04 | 170.26 ± 0.11 |
| | <200 nm | 184 | 0.13 | -0.66 ± 0.05 | 38.60 ± 0.12 |
| | 200-800 nm | 450 | 0.18 | -0.98 ± 0.07 | 11.90 ± 0.19 |
| | >800 nm | 812 | 0.59 | -0.44 ± 0.05 | 9.84 ± 0.12 |
| Murchison residues | Colloidal | 34 | 0.09 | -1.77 ± 0.06 | 148.03 ± 0.17 |
| | <200 nm | 155 | 0.18 | -0.43 ± 0.08 | 83.78 ± 0.18 |
| | 200-800 nm | 350 | 0.33 | -0.06 ± 0.07 | 6.78 ± 0.17 |
| | >800 nm | 1,437 | 0.26 | 0.02 ± 0.03 | 4.96 ± 0.08 |

**Notes.**

$\varepsilon^i$Cr=[$(^i$Cr/$^{52}$Cr$)_{sample}$/$(^i$Cr/$^{52}$Cr$)_{SRM3112a}$-1]×10$^4$. Mass fractionation was corrected by internal normalization to a fixed $^{50}$Cr/$^{52}$Cr ratio of 0.051859 using the exponential law. Uncertainties are 95 % confidence intervals.

**Table 4**
Cr and Fe isotopic analyses of $^{54}$Cr-rich spots

|  | $\varepsilon^{53}$Cr | $\varepsilon^{54}$Cr | $\varepsilon^{57}$Fe |
|---|---|---|---|
| np1 | -1 ± 288 | 25,861 ± 2,903 | -929 ± 1,281 |
| np1 replicate | -29 ± 447 | 17,902 ± 3,291 | 2,578 ± 2,409 |
| np2 | -59 ± 253 | 2,654 ± 915 | 917 ± 1,644 |

**Notes.**
$\varepsilon^i$Cr=[($^i$Cr/$^{52}$Cr)$_{ROI}$/($^i$Cr/$^{52}$Cr)$_{bulk\ image}$-1]×10$^4$, $\varepsilon^{57}$Fe=[($^{57}$Fe/$^{56}$Fe)$_{ROI}$/($^{57}$Fe/$^{56}$Fe)$_{bulk\ image}$-1]×10$^4$. Uncertainties are 95 % confidence intervals. Note that contrary to TIMS data (Table 3), $\varepsilon$-values have not been corrected for mass fractionation by internal normalization.